\documentclass{article}
\usepackage{arxiv}

\usepackage[utf8]{inputenc} 
\usepackage[T1]{fontenc}    
\usepackage{hyperref}       
\usepackage{url}            
\usepackage{booktabs}       
\usepackage{amsfonts}       
\usepackage{nicefrac}       
\usepackage{microtype}      
\usepackage{lipsum}
\usepackage{graphicx}
\usepackage{amsmath}
\graphicspath{{images/}} 
\usepackage{wasysym} 

\usepackage{longtable}
\usepackage{comment}
\usepackage{tikz}
\tikzstyle{arrow} = [thick,->,>=stealth]

\usepackage{forest}

\usetikzlibrary{trees,positioning,shapes,shadows,arrows.meta}

\title{Security in LLM-as-a-Judge: A Comprehensive SoK}

\author{
\textbf{Aiman Al Masoud}\textsuperscript{1}, \textbf{Antony Anju}\textsuperscript{2}, \textbf{Marco Arazzi}\textsuperscript{1}, \textbf{Mert Cihangiroglu}\textsuperscript{1}, \textbf{Vignesh Kumar Kembu}\textsuperscript{1}, \\
\textbf{Serena Nicolazzo}\textsuperscript{1*}, \textbf{Antonino Nocera}\textsuperscript{1}, \textbf{Vinod P.}\textsuperscript{2}, \textbf{Saraga Sakthidharan}\textsuperscript{1} \\
\\
\textsuperscript{1}Department of Electrical, Computer and Biomedical Engineering, University of Pavia, Pavia, Italy \\
\textsuperscript{2}Department of Computer Applications, Cochin University of Science and Technology, Kerala, India \\
\textsuperscript{*}Corresponding author
}

\begin{document}
\maketitle

\begin{abstract}
LLM-as-a-Judge (LaaJ) is a novel paradigm in which powerful language models are used to assess the quality, safety, or correctness of generated outputs. While this paradigm has significantly improved the scalability and efficiency of evaluation processes, it also introduces novel security risks and reliability concerns that remain largely unexplored. In particular, LLM-based judges can become both targets of adversarial manipulation and instruments through which attacks are conducted, potentially compromising the trustworthiness of evaluation pipelines. In this paper, we present the first Systematization of Knowledge (SoK) focusing on the security aspects of LLM-as-a-Judge systems. We perform a comprehensive literature review across major academic databases, analyzing 863 works and selecting 45 relevant studies published between 2020 and 2026. Based on this study, we propose a taxonomy that organizes recent research according to the role played by LLM-as-a-Judge in the security landscape, distinguishing between attacks targeting LaaJ systems, attacks performed through LaaJ, defenses leveraging LaaJ for security purposes, and applications where LaaJ is used as an evaluation strategy in security-related domains. We further provide a comparative analysis of existing approaches, highlighting current limitations, emerging threats, and open research challenges. Our findings reveal significant vulnerabilities in LLM-based evaluation frameworks, as well as promising directions for improving their robustness and reliability. Finally, we outline key research opportunities that can guide the development of more secure and trustworthy LLM-as-a-Judge systems.
\end{abstract}

\keywords{Large Language Model \and Security \and LLM Judgment \and LLM Evaluation \and LaaJ \and LLM attack.}

\section{Introduction}

In recent years, Large Language Models (LLMs) have transformed the landscape of artificial intelligence by enabling advanced capabilities in text generation, reasoning, and decision support \cite{zhao2026survey,thirunavukarasu2023large,kasneci2023chatgpt}. Currently, a new paradigm known as LLM-as-a-Judge (LaaJ) has emerged, in which powerful language models are employed to evaluate the outputs of other models or systems. By leveraging their strong language understanding and reasoning capabilities, LLM-based judges can assess aspects such as correctness, quality, safety, and alignment of generated content. This paradigm has quickly gained popularity in benchmarking frameworks, reinforcement learning pipelines, automated evaluation systems, and AI-assisted review processes.

While the benefits of LaaJ have been widely recognized, its integration into critical evaluation pipelines also raises important security concerns. Empirical studies have highlighted inherent reliability issues in LLM-as-a-Judge systems, such as position bias, where evaluation outcomes are influenced by the order of candidate solutions rather than their true quality \cite{shi2025judging} without providing objective, systematic, replicable, and deterministic evaluation. In particular, LLM-based judges can become targets of adversarial manipulation, may be exploited as tools to facilitate attacks, or can be leveraged as mechanisms to support defensive analysis and security evaluation.

Despite the growing number of studies addressing these aspects, the literature remains fragmented, with contributions scattered across different research communities such as natural language processing, software security, and AI safety. As a result, there is currently a lack of a structured framework that systematically organizes the emerging research on the security implications of the LaaJ paradigm.

In this study, we performed a comprehensive literature search on the topics described above. Our review primarily focused on journal and conference publications from recent years, retrieved through the most important databases, such as Google Scholar\footnote{https://scholar.google.com}, IEEE Xplore\footnote{https://ieeexplore.ieee.org}, ACM Digital Library\footnote{https://dl.acm.org}, Springer\footnote{https://link.springer.com}, and ScienceDirect\footnote{https://www.sciencedirect.com}. To identify relevant works, we conducted the search using combinations of the following keywords: \textit{``LLM-as-a-Judge''} together with security-related terms such as \textit{``security''}, \textit{``privacy''}, \textit{``toxicity''}, and \textit{``reliability''}. Selection criteria prioritized high-quality sources, specifically Q1 journals and A*/A-ranked conference papers. Duplicate entries, literature reviews, surveys, and non-English publications were excluded. Given the novelty of the LLM-as-a-Judge research area, we also considered a limited number of relevant preprints available on arXiv. These works were included to capture emerging contributions that have not yet completed the peer-review process but address important aspects of the topic. To ensure a quality threshold, we selected only preprints in which at least one author had a well-established academic profile, defined as having more than $5,000$ citations. In total, 863 works were analyzed, of which 45 papers published between 2020 and 2026 were selected for detailed review.

Our systematic analysis aims to uncover existing gaps and emerging trends in the security landscape of LLM-as-a-Judge (LaaJ) systems. By structuring and critically examining the current literature, we aim to guide researchers toward improving the robustness and reliability of LaaJ-based systems and to stimulate the development of more secure evaluation frameworks.

The main contributions of this paper are:

\begin{itemize}
\item We provide the first comprehensive overview of the security aspects of LLM-as-a-Judge systems, covering attacks targeting LaaJ, attacks performed through LaaJ, defenses leveraging LaaJ, and their use as evaluation mechanisms in security applications.

\item We propose a novel taxonomy that systematically categorizes the existing literature according to the role played by LaaJ in the security landscape, distinguishing between LaaJ as an attack target, as an attack instrument, as a defense tool, and as an evaluation strategy.

\item We conduct a comparative analysis of the selected studies, highlighting current limitations, research gaps, and emerging trends in LaaJ security.

\item We discuss open challenges and outline promising future research directions to improve the robustness, reliability, and trustworthiness of LLM-as-a-Judge systems.
\end{itemize}

The remainder of this paper is organized as follows. Section \ref{sec:background} provides the essential background on LLM-as-a-Judge, including commonly used evaluation metrics and existing benchmarks. Section \ref{sec:related} reviews the most relevant surveys and related studies in this emerging research area. Section \ref{sec:taxonomy} introduces the taxonomy adopted in this work to classify the security-related literature on LLM-as-a-Judge. Based on this taxonomy, Section \ref{sec:attack} analyzes the threat landscape, discussing attacks targeting LLM-as-a-Judge systems as well as attacks performed through them. Section \ref{sec:defense} presents approaches that leverage LLM-as-a-Judge as a defense mechanism or security analysis tool. Section \ref{sec:evaluation} examines the use of LLM-as-a-Judge as an evaluation strategy in security-related applications. Section \ref{sec:challenge} discusses open challenges, limitations, and promising research directions. Finally, Section \ref{sec:conclusion} concludes the paper.

\section{Definitions and Background}
\label{sec:background}
This section presents the essential background necessary to understand the key concepts discussed in this paper. In particular, it provides a concise overview of the main definition and key concepts related to LLM-as-a-Judge. Furthermore, Table \ref{tab:SystemSymbols} summarizes the acronyms and notation adopted throughout the paper for ease of reference.

\begin{table}[ht]
\centering
  \caption{Summary of the acronyms used in the paper}
  \begin{tabular}{ll}
\hline
    \textbf{Symbol} & \textbf{Description}\\
\hline
    AA & Adversarial Attack\\
    CTI & Cyber Threat Intelligence\\
    DL & Deep Learning\\
    FL & Federated Learning\\
    GPT & Generative Pre-trained Transformer\\
    ICL & In-Context Learning\\
    IoT & Internet of Things\\
    LaaJ & LLM-as-a-Judge \\
    LLM & Large Language Model\\
    ML & Machine Learning\\
    RAG & Retrieval-Augmented Generation\\
\hline
\end{tabular}
\label{tab:SystemSymbols}
\end{table}

\subsection{Definition of LLM-as-a-Judge}
An LLM-as-a-Judge is defined as a function implemented by a Large Language Model (LLM) that produces an evaluation outcome for one or more input samples conditioned on a given evaluation context:

$$R = P_{\theta}(X_n, C)$$
\noindent
where:

\begin{itemize}
    \item $\mathbf{P_{\theta}}$ denotes the evaluation function induced by a LLM parameterized by $\theta$. The model can be either a general-purpose foundation model or a fine-tuned variant, and generates outputs through an autoregressive process.

    \item $\mathbf{X_n = \{x_1, \dots, x_n\}}$ represents the set of samples to be evaluated. Inputs may belong to different modalities, including text, images, code, or multimodal data. The value of $n$ determines the evaluation setting:
    \begin{itemize}
        \item $n = 1$: \emph{point-wise evaluation}, producing an absolute score or assessment;
        \item $n = 2$: \emph{pair-wise evaluation}, producing a comparative judgment;
        \item $n > 2$: \emph{list-wise evaluation}, producing a ranking over multiple candidates.
    \end{itemize}

    \item $\mathbf{C}$ denotes the evaluation context, which may include task instructions, evaluation criteria, demonstration examples, prompt templates, or interaction history.

    \item $\mathbf{R}$ is the evaluation output generated by the LLM-as-a-Judge, which may consist of a numerical score, label, ranking, preference decision, or natural language rationale, either absolute or relative. In general, we distinguish three main types of judgment outputs. The first one is a score-based protocol when each candidate sample is assigned a continuous or discrete score, i.e., $R = \{C_1 : S_1, \dots, C_n : S_n\}$. Alternatively, the judgment may take the form of a ranking, where candidates are ordered according to their relative quality, expressed as $R = \{C_i \succ \dots \succ C_j\}$. Finally, the judgment can be selection-based, where the LLM identifies one or more optimal candidates from the set, represented as $R = \{C_i, \dots, C_j\} \succ \{C_1, \dots, C_n\}$.

\end{itemize}

\begin{figure*}
    \centering
    \includegraphics[width=0.65\linewidth]{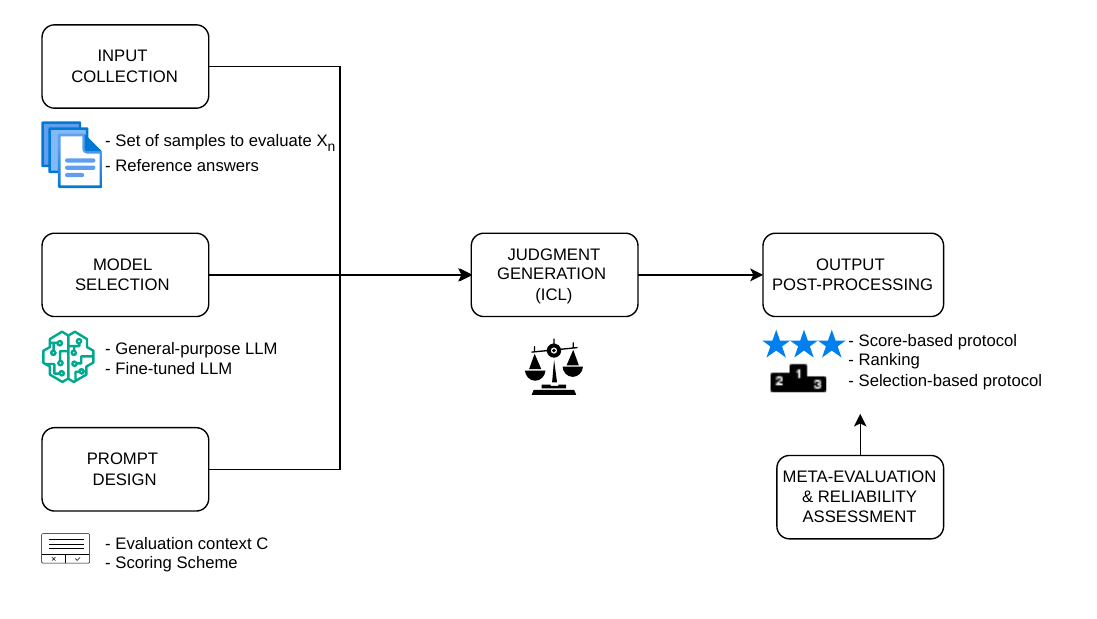}
    \caption{General workflow for LLM-as-a-Judge scheme}
    \label{fig:scheme}
\end{figure*}

A basic LLM-as-a-Judge system typically follows a structured workflow consisting of several key steps, also visible in Figure \ref{fig:scheme}:

\begin{enumerate}
    \item \textbf{In-Context Learning (ICL)}. ICL \cite{dong2024survey} can be described as conditional generation over an augmented context containing demonstration examples, without parameter updates. The evaluation task is specified directly within the prompt through instructions and, optionally, a small number of annotated examples. The model does not update its parameters; instead, it infers the evaluation pattern from contextual cues provided at inference time. In other words. More formally, given an input object 
    $x$ to evaluate (text, image, code, etc.); a set of candidate evaluations $Y = \{y_1, \dots, y_m\}$; a demonstration set (or context) $C = \{ I, s(x_1, y_1), \dots, s(x_k, y_k) \}$, where $I$ is an optional instruction, and $s(x_i, y_i)$ are demonstration pairs showing input and evaluation; a pretrained LLM $M$ assigns a score to each candidate evaluation:

    $$S_\theta(y_j \mid x, C) = f_M(y_j, x, C)$$

    \noindent
    where $f_M$ is the scoring function induced by the autoregressive model over the concatenated input and context. If a score-based protocol is applied, the LLM-as-a-Judge selects the highest-scoring candidate evaluation:

   $$ \hat{y} = \arg\max_{y_j \in Y} S_\theta(y_j \mid x, C)$$

    \noindent meaning that the candidate with the highest score is chosen as the final evaluation $\hat{y}$.
    \item \textbf{Model Selection}. The choice of the underlying LLM significantly affects evaluation quality. Factors such as model size, alignment strategy, domain specialization, and reasoning capability influence consistency, bias, and robustness. One may choose either a general-purpose or a fine-tuned LLM. General LLMs provide broad flexibility, while fine-tuned LLMs are more reliable in specific domains. In both cases, the evaluation is performed via in-context learning, so the model parameters remain fixed at inference.
    \item \textbf{Prompt Design}. Careful prompt engineering defines the evaluation criteria, output format, and constraints. The prompt may include scoring rubrics (how each evaluation criterion is measured), comparison instructions, or structured templates to reduce ambiguity and variability in the model's judgments.
    \item \textbf{Output Post-Processing}. The raw model output is often transformed into structured and usable formats. This may include extracting numerical scores, normalizing labels, aggregating multiple runs, or applying calibration techniques to reduce randomness.
    \item \textbf{Evaluation Pipeline}. After finalizing the above processes, we obtain the final evaluation. At this point, the system can incorporate validation procedures to assess performance and reliability. This can involve benchmarking against human judgments, measuring inter-run consistency, analyzing bias, and testing robustness under input perturbations.
\end{enumerate}

LLMs can act as evaluators across a variety of tasks, providing scalable and flexible judgment where human assessment is costly or impractical. Their applications span models, data, autonomous agents, and reasoning processes, each requiring tailored evaluation strategies \cite{gu2024survey}. In particular:

\begin{itemize}
    \item LLM-as-a-Judge for models evaluates the outputs of other machine learning or language models, such as ranking responses, scoring generated content, or assessing model performance on specific tasks.
    \item LLM-as-a-Judge for data assesses the quality, relevance, or correctness of data samples, including labeling, filtering noisy inputs, or verifying annotations.
    \item LLM-as-a-Judge for agents monitors and evaluates the actions or decisions of autonomous agents, providing feedback, ranking strategies, or verifying compliance with desired behavior.
    \item LLM-as-a-Judge for reasoning or thinking evaluates logical reasoning, problem-solving steps, or decision-making processes, checking consistency, coherence, and alignment with human-like reasoning.
\end{itemize}

\subsection{Evaluation of LLM-as-a-Judge}
Meta-evaluation is the process of assessing the quality of the evaluator itself, and it is essential for determining the consistency, reliability, and validity of LLM judges. As LLMs are used in many different evaluation tasks, meta-evaluation methods have also evolved. Researchers have introduced various datasets and metrics tailored to specific goals to measure how trustworthy these evaluations are \cite{li2024llmsasjudges}.

\begin{itemize}
    \item \textbf{Benchmark}: A common way to evaluate LLM-based judges is by measuring how well their judgments align with human preferences, which are often treated as the gold standard. Due to the wide variety of LLM applications, different benchmarks have been developed for specific evaluation goals. 
    \item \textbf{Metrics}: Evaluating LLMs-as-Judges focuses on how closely the model's judgments match human evaluations. High agreement with human ratings is a key sign of performance, especially given the complexity and subjectivity of many tasks.
\end{itemize}

Section \ref{subsec:targetEval} presents the papers that deal with LaaJ as the target of evaluation with the considered benchmarks and metrics used.

\section{Related Work}
\label{sec:related}

This section reviews existing literature while identifying the gap in security-focused analyses. Indeed, most of the recent surveys on LLM-as-a-Judge focus on general aspects of this emerging paradigm, providing overviews of its functionality, methodology, applications, and limitations \cite{gu2024survey,li2024llmsasjudges}. However, dedicated studies examining security risks, adversarial vulnerabilities, and the robustness of LLM-based evaluation systems remain scarce.

In particular, the survey \cite{gu2024survey} provides a formal definition and taxonomy of the LLM-as-a-Judge paradigm. It focuses on building reliable LLM-based evaluation systems by proposing strategies for improving consistency and bias mitigation, along with a novel reliability-oriented benchmark.
The authors of \cite{li2024llmsasjudges} explore the emerging paradigm of LLMs-as-Judges, analyzing their functionality, methodology, applications, meta-evaluation, and limitations. Moreover, the include a brief paragraph on adversarial attacks on LLMs-as-Judges within the limitations section. Analogously, the survey \cite{li2025generation} proposes a systematic taxonomy along three dimensions, namely what to judge (attributes), how to judge (methodology), and how to benchmark (evaluation). In this paper, security aspects are briefly considered as one of the evaluation outputs or attributes assessed by the LLM-as-a-Judge, alongside other qualities such as helpfulness, reliability, and relevance.

A limited number of surveys analyze specific application domains of LLM-as-a-Judge \cite{he2025llm,genovese2026artificial,kocaman2025clinical,nawara2025comprehensive}. For instance, \cite{he2025llm} focuses on software engineering and explores how LLMs can automatically evaluate code and software artifacts, highlighting the limitations of human and traditional automated evaluation, proposing a roadmap to develop reliable, scalable, and nuanced LLM-based evaluation frameworks by 2030. In this paper, adversarial threats are mentioned in the limitations section, but the discussion is limited to the specific application domain of software engineering. Whereas \cite{genovese2026artificial,kocaman2025clinical} provide a narrative review of LLM-as-a-Judge in healthcare. In particular, \cite{genovese2026artificial} focuses on applications such as clinical documentation, medical Q\&A, and conversation assessment. It finds that LLM judges align well with clinicians on objective criteria (e.g., factuality, grammar, internal consistency) and can approach inter-clinician agreement, especially when using structured evaluation and chain-of-thought prompting. The authors of \cite{kocaman2025clinical} review lots of existing evaluation methods, including multiple-choice questions, log-likelihood, perplexity, and also LLMs-as-a-Judges. Finally, the survey in \cite{nawara2025comprehensive} reviews LLM-based recommendation systems, analyzing different architectures (discriminative, generative, hybrid, graph-enhanced, and multimodal) and techniques such as fine-tuning, prompt engineering, and retrieval-augmented generation. It also discusses LLM-as-a-Judge as a possible evaluation approach.

Other surveys having a precise focus are presented in \cite{you2026agent,yu2025aisjudge}. Specifically, they concentrate on the methodological evolution of LLM-as-a-Judge toward agentic evaluation frameworks, called Agent-as-a-Judge, introducing a unified framework and taxonomy for agentic evaluation systems. The survey \cite{cao2025toward} frames LLM-as-a-Judge within the broader transition from manual to automated evaluation of LLMs. However, it treats LLM-based judging as one component of the evaluation pipeline rather than as a standalone paradigm, and it does not provide a dedicated analysis of security risks or adversarial vulnerabilities in LLM-as-a-Judge systems.

\begin{table*}
\caption{Survey papers related to our work}
\scriptsize
\centering
\begin{tabular}{p{1.8cm}|p{1cm}p{1.2cm}p{1cm}p{1.2cm}p{1.2cm}p{2.5cm}p{3.5cm}}
    \hline
    \textbf{Paper} & \textbf{Year} & \textbf{Timeline} & \textbf{\# Papers} & \textbf{Definition} & \textbf{Evaluation} & \textbf{Security Aspects} &  \textbf{Domain/Focus} \\
    \hline
    
    Gu et al. \cite{gu2024survey} & 2024 & 2023-2025 & $\sim$ 220 & \CIRCLE & \CIRCLE & \Circle & NLP, Social Intelligence, Multi-Modal Evaluation, and others\\

    Xu et al. \cite{xu2024benchmark} & 2024 & 2020-2024 & $\sim$ 180 & \Circle & \LEFTcircle & \LEFTcircle (Security aspects related to system trustworthiness) & Benchmark Data Contamination (BDC) \\

    Li et al. \cite{li2025generation} & 2024 & 2022-2025 & $\sim$ 270 & \CIRCLE & \CIRCLE & \LEFTcircle (Briefly mentioned as evaluation outputs) & Evaluation, Alignment, Retrieval, and Reasoning \\
        
    Li et al. \cite{li2024llmsasjudges} & 2024 & 2023-2024 & $\sim$ 300 & \CIRCLE & \CIRCLE & \LEFTcircle (Adversarial Attack briefly mentioned) & NLP, Multi-Modal Evaluation, Information retrieval, and others\\

    Cao et al. \cite{cao2025toward} & 2025 & 2023-2025 & $\sim$ 370 & \LEFTcircle & \CIRCLE & \Circle & - \\

    He et al. \cite{he2025llm} & 2025 & 2023-2025 & 42 & \CIRCLE & \CIRCLE & \LEFTcircle (Security aspects related to system trustworthiness) & Software Engineering\\

    Kocaman et al. \cite{kocaman2025clinical} & 2025 & 2023-2025 & $\sim$ 30 & \LEFTcircle & \LEFTcircle & \LEFTcircle (Security aspects related to system trustworthiness) & Healthcare \\

    Li et al. \cite{li2025software} & 2025 & 2020-2024 & $\sim$ 50 & \CIRCLE & \CIRCLE & \LEFTcircle (Briefly mentioned as limitation) & Software Engineering\\
    
    Nawara et al. \cite{nawara2025comprehensive} & 2025 & 2020-2025 & $\sim$ 150 & \LEFTcircle & \LEFTcircle & \LEFTcircle (Security aspects related to system trustworthiness) & Recommandation System \\

    Yu et al. \cite{yu2025aisjudge} & 2025 & 2023-2025 & $\sim$ 60 & \LEFTcircle & \LEFTcircle & \Circle & Agent-as-a-Judge \\

    Genovese et al. \cite{genovese2026artificial} & 2026 & 2023-2025 & $\sim$ 65 & \CIRCLE & \CIRCLE & \Circle & Healthcare \\
        
    You et al. \cite{you2026agent} & 2026 & 2023-2025 & $\sim$ 40 & \LEFTcircle & \LEFTcircle & \Circle & Agent-as-a-Judge \\

    \hline
    \textbf{Our Survey} & \textbf{2026} & \textbf{2020-2026} & \textbf{45} & \CIRCLE & \CIRCLE & \CIRCLE & \textbf{Cross-domain (Comprehensive)} \\
    \hline
\end{tabular}
\label{tab:relatedSurveys}
\end{table*}

\begin{figure*}[ht]
    \centering
    
\tikzset{
    basic/.style  = {draw, text width=3cm, align=center, fill=gray!60, font=\sffamily, rectangle, rounded corners=2pt,text width=2.5cm},
    onode/.style = {basic, thin, rounded corners=2pt, align=left, fill=gray!10, text width=3cm},
    tnode/.style = {basic, align=left, fill=white, text width=4cm},
    xnode/.style = {basic, thin, rounded corners=2pt, align=center, fill=gray!30}
}

\begin{forest} for tree={
    grow=east,
    growth parent anchor=west,
    parent anchor=east,
    child anchor=west,
    edge path={\noexpand\path[\forestoption{edge},->, >={latex}] 
         (!u.parent anchor) -- +(5pt,0pt) |-  (.child anchor) 
         \forestoption{edge label};}
}
[LaaJ\\Classification, basic,  l sep=10mm,
    [Evaluation, xnode,  l sep=6mm,
        [LaaJ as tool \cite{deldjoo2025toward,zhang2025lsrp,cantini2025benchmarking,li2025lexrag,zhou2025rescriber,wang2025manipulating,balog2025rankers,abeyratne2025alignllm,shen2025defining,xu2025towards,stavarache2025enhancing,singh2026multi,krayko2025rurage,goldman2025types,olewicki2026impact,wang2025multi,sghaier2025harnessing,jaoua2025combining,baek2026llm,ji2026codegen,zhou2025se,wang2026fine,morales2025impromptu,shi2025humanin,zahan2025leveraging,webb2025synthetic,belcastro2025enhancing,blefari2025cyberrag,farrukh2025xg,pasini2026evaluating,shao2025effective}, tnode]
        [LaaJ as target \cite{ye2024justice,lai2026bias}, tnode]
    ]
    [Defense, xnode,  l sep=6mm,
        [LaaJ as Mitigation Strategy \cite{mao2025towards,li2025whos,shen2025gptracker}, tnode]
    ]
    [Attack, xnode,  l sep=6mm,
        [LaaJ as tool \cite{liu2025compromising}, tnode]
        [LaaJ as target \cite{raina2024llm,shi2024optimization,wei2025emoji,li2025llmsreliablyjudgeyet,maloyan2025adversarial,tong2025bad,zhao2025tokenfool,ding2026rubrics} , tnode]
    ]
]
\end{forest}

    \caption{Literature Survey Tree}
    \label{fig:classification}
\end{figure*}
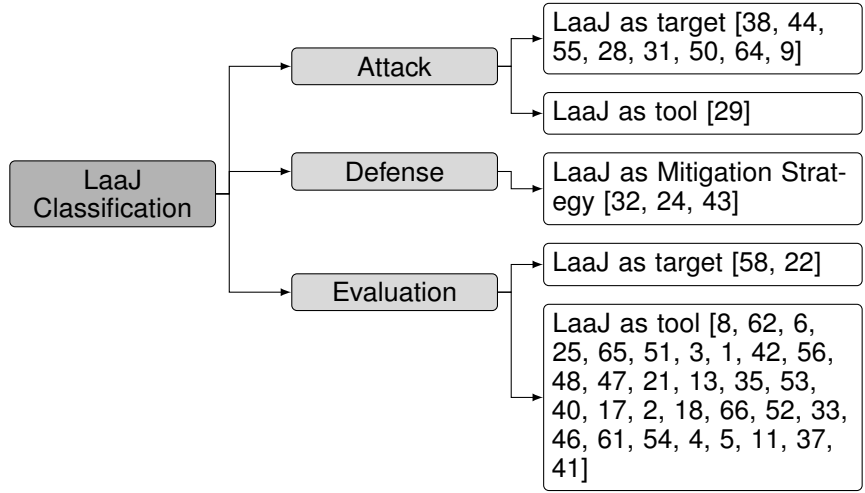

This analysis underscores a clear research gap: while LaaJ has been widely studied, security-focused surveys covering adversarial robustness, attack vectors, and defensive strategies are largely missing. Our survey addresses this gap by systematically analyzing the security aspects of LaaJ systems across different scenarios, thereby offering a unique and timely contribution to the field. Table \ref{tab:relatedSurveys} presents a comparative analysis of existing surveys in relation to our work. Specifically, the table highlights: {\em (i)} publication year, {\em (ii)} literature timeline, {\em (iii)} number of papers included in the survey, {\em (iv)} whether a formal definition of LaaJ is provided, {\em (v)} whether the survey treats evaluation as a systematic measurement of outputs via LaaJ systems, {\em (vi)} whether security aspects of LaaJ are addressed, and {\em (vii)} the application domain or specific focus of the survey.

\section{Taxonomy of LLM-as-a-Judge in the security context}
\label{sec:taxonomy}
In this section, we present the taxonomy that categorizes existing works according to the role played by LaaJ in the security landscape. Specifically, our taxonomy distinguishes five main perspectives: {\em(i)} attacks targeting LaaJ systems, where the goal is to manipulate the behavior of the judge; {\em(ii)} attacks performed through LaaJ, where LLM-based judges are exploited as instruments to conduct malicious activities; {\em(iii)} defenses leveraging LaaJ, where LLM-based evaluators are used to detect security threats; {\em(iv)} papers evaluating LaaJ; and {\em(v)} applications where LaaJ is adopted as an evaluation strategy in security-related tasks. This taxonomy enables a clearer understanding of how LaaJ interacts with security mechanisms and threats. 

Figure \ref{fig:classification} represents our adopted classification, whereas Table \ref{tab:purpose} presents a summary of all the selected papers, including the year of publication, the main scope of the paper, the classification, and the application context of the contribution.

{
\scriptsize
\begin{longtable}{p{1.8cm}lp{5.5cm}p{1.5cm}p{4.5cm}}
\caption{Summary of the main purpose of all the analysed papers\label{tab:purpose}}\\

\hline
    \textbf{Ref.} & \textbf{Year} & \textbf{Scope} & \textbf{Taxonomy} & \textbf{Context}\\
\hline
    Raina et al. \cite{raina2024llm} & 2024 & Adversarial robustness of LLMs used as assessors & Attack & Assessment and benchmarking using LLMs\\
    
    Shi et al. \cite{shi2024optimization} & 2024 & Prompt injection attack that manipulates LaaJ systems to select an attacker-chosen response & Attack & Decision-making systems using LaaJ\\

    Ye et al. \cite{ye2024justice} & 2024 & Evaluate reliability of LaaJ & Evaluation & - \\
    
    Wei et al. \cite{wei2025emoji} & 2025 & Introduces Emoji Attack that exploits token segmentation bias to trick LaaJ & Attack & -\\
        
    Abeyratne et al. \cite{abeyratne2025alignllm} & 2025 & Propose an unsupervised ensemble method to evaluate LLM-generated responses & Evaluation & Law\\

    Balog et al. \cite{balog2025rankers} & 2025 & Empirical study that investigates biases in LaaJ within information retrieval pipelines & Evaluation & Information Retrieval \\

    Belcastro et al. \cite{belcastro2025enhancing} & 2025 & Evaluate structured, human-readable reports about detected threats & Evaluation & CTI \\

    Cantini et al. \cite{cantini2025benchmarking} & 2025 & Proposing an evaluation tool using LaaJ to assess adversarial bias robustness & Evaluation & General \\

    Deldjoo et al. \cite{deldjoo2025toward} & 2025 & Evaluating generative recommender systems (Gen-RecSys) & Evaluation & Recommender systems\\

    Krayko et al. \cite{krayko2025rurage} & 2025 & Propose an open-source framework for evaluating QA responses & Evaluation & Industrial NLP systems\\
    
    Li et al. \cite{li2025lexrag} & 2025 & Assess the quality and reliability of a benchmark designed to evaluate RAG systems & Evaluation & Multi-turn legal consultations\\ 

    Li et al. \cite{li2025llmsreliablyjudgeyet} & 2025 & Evaluate the robustness of LaaJ systems against adversarial attacks, prompt variations, and model choices & Attack & Web platforms\\

    Maloyan et al. \cite{maloyan2025adversarial} & 2025 & Evaluate the vulnerability of LaaJ systems to prompt injection attacks & Attack & -\\
    
    Mao et al. \cite{mao2025towards} & 2025 & Software vulnerability detection and explanation & Defense & Software development\\

    Farrukh et al. \cite{farrukh2025xg} & 2025 & Evaluation explanation generation and suggesting mitigations for real-time intrusion detection & Evaluation & Network Intrusion Detection Systems\\

    Goldman et al. \cite{goldman2025types} & 2025 & Evaluation of large code review datasets & Evaluation & Software development / Code review\\

    Jaoua et al. \cite{jaoua2025combining} & 2025 & Automatic evaluation and generation of code review comments & Evaluation & Software development / Code review\\

    Li et al. \cite{li2025whos} & 2025 & Introduces judgment detection, a method to identify when scores are generated by a LaaJ & Defense & -\\

    Liu et al. \cite{liu2025compromising} & 2025 & Introduces Contextual Backdoor Attacks performed also through LaaJ & Attack & - \\
    
    Morales et al. \cite{morales2025impromptu} & 2025 & A framework for designing, managing, and reusing prompts for generative AI & Evaluation & Software development / Code review\\

    Sghaier et al. \cite{sghaier2025harnessing} & 2025 & Automatic code review comments generated by LLM evaluation & Evaluation & Software development / Code review\\

    Shao et al. \cite{shao2025effective} & 2025 & Automated assessment of LLM agents performing cybersecurity attacks & Evaluation & Attack detection\\
    
    Shen et al. \cite{shen2025defining} & 2025 & Risk-aware evaluation framework for LLM & Evaluation & Healthcare and Finance\\

    Shen et al. \cite{shen2025gptracker} & 2025 & Evaluate GPT behaviors and identify misuses & Defense & GPT ecosystem\\

    Shi et al. \cite{shi2025humanin} & 2025 & Evaluation of the validity of program patches & Evaluation & Software development / Code review\\
    
    Stavarache \cite{stavarache2025enhancing} & 2025 & Evaluation of the alignment of LLMs & Evaluation & Banking Industry Architecture Network\\

    Tong et al. \cite{tong2025bad} & 2025 & Introduces a Backdoor threat targeting LaaJ systems & Attack & -\\

    Xu et al. \cite{xu2025towards} & 2025 & Evaluation benchmarks to assess reliability and quality of LLM outputs & Evaluation & Water and wastewater management\\

    Wang et al. \cite{wang2025manipulating} & 2025 & Evaluate a cross-modal prompt injection attack & Evaluation & Multimodal autonomous agents\\

    Wang et al. \cite{wang2025multi} & 2025 & Automatically evaluate and rank testing crowdsourced testing reports using LLMs & Evaluation & Software development / Code review\\

    Webb et al. \cite{webb2025synthetic} & 2025 & Evaluate the quality of the generated social engineering scenarios & Evaluation & Cybersecurity training\\
    
    Zahan et al. \cite{zahan2025leveraging} & 2025 & Analyze code and classify packages as malicious or benign & Evaluation & Malware Detection \\

    Zhang et al. \cite{zhang2025lsrp} & 2025 & Evaluation of privacy-preserving cloud-device collaboration & Evaluation & Cloud Computing \\

    Zhao et al. \cite{zhao2025tokenfool} & 2025 & Investigate a vulnerability in LaaJ systems used as reward models & Attack & -\\
    
    Zhou et al . \cite{zhou2025rescriber} & 2025 & Evaluation of privacy-preserving cloud-device collaboration & Evaluation & Software development / Code review \\

    Zhou et al. \cite{zhou2025se} & 2025 & Propose an LLM-as-Ensemble-Judge evaluation metric for assessing the correctness of software artifacts generated by LLMs & Evaluation & Chatbot \\
    
    Baek et al. \cite{baek2026llm} & 2026 & Evaluate whether code was AI-generated & Evaluation & Software development / Code review\\

    Blefari et al. \cite{blefari2025cyberrag} & 2026 & Evaluate and justify the threat classification &  Evaluation & Malware / Attack Detection \\

    Ding et al. \cite{ding2026rubrics} & 2026 & Identifies Rubric-Induced Preference Drift attack & Attack & -\\
    
    Ji et al. \cite{ji2026codegen} & 2026 & Present a benchmark for evaluating text-to-3D modeling via code generation from LLMs & Evaluation & Software development / Code review\\

    Lai et al. \cite{lai2026bias} & 2026 & Discover unknown biases in LaaJ evaluation systems & Evaluation & - \\
    
    Olewicki et al. \cite{olewicki2026impact} & 2026 & Automatic code review comments generated by LLM evaluation & Evaluation & Software development / Code review\\

    Pasini et al. \cite{pasini2026evaluating} & 2026 & Generating security-aware functions (attack detectors) in software development &  Evaluation & Malware / Attack Detection \\
    
    Singh et al. \cite{singh2026multi} & 2026 & Evaluate construction schedules and task descriptions & Evaluation & Architecture, Engineering, and Construction\\
    
    Wang et al. \cite{wang2026fine} & 2026 & Assess the outputs of static and dynamic analyses to determine whether code slices are vulnerable & Evaluation & Software development / Code review\\
\hline    
\end{longtable}
}

\section{Threat Landscape for LLM-as-a-Judge}
\label{sec:attack}

As LLM-as-a-Judge systems are increasingly deployed in high-stakes settings, from model benchmarking and safety filtering to reinforcement learning pipelines, understanding their security vulnerabilities becomes essential. This section surveys the threat landscape from two complementary perspectives. First, we examine scenarios where the Judge itself is the target of adversarial attacks, covering both training-time and inference-time threat vectors \cite{raina2024llm, shi2024optimization, wei2025emoji, li2025llmsreliablyjudgeyet, maloyan2025adversarial, tong2025bad, zhao2025tokenfool, ding2026rubrics}. Second, we examine scenarios where a Judge LLM is employed as an instrument to facilitate attacks against other systems \cite{liu2025compromising}.

Table \ref{tab:threat-summary} illustrates a structured survey of adversarial attacks targeting LLM-as-a-Judge systems as well as cases where LLM-as-a-Judge is used as an attack instrument, cataloguing each study by its publication year, the role the judge model plays in the pipeline, and the category and specific type of attack employed. For each entry, it records which LLM serves as the judge, the quantitative metrics used to measure attack effectiveness, the headline results demonstrating how severely the attack compromises judge behaviour, and, where applicable, the benchmark or dataset on which the attack was assessed.

\begin{table*}[ht]
\caption{Summary of Threat Landscape for LLM-as-a-Judge}
\label{tab:threat-summary}
\scriptsize
\centering
\resizebox{\textwidth}{!}{
\begin{tabular}{p{1cm}p{.4cm}p{1.6cm}p{1.5cm}p{2cm}p{2.3cm}p{2cm}p{2.5cm}p{2cm}}
\hline
\textbf{Ref} & \textbf{Year} & \textbf{Role of the Judge} & \textbf{Attack Category} & \textbf{Attack Type} & \textbf{Judge Model(s)} & \textbf{Eval Metrics} & \textbf{Main Results} & \textbf{Benchmark} \\
\hline

Raina et al. \cite{raina2024llm} & 2024 & Zero-shot assessor for text quality & Universal adversarial attack & Short concatenated phrases (2--5 tokens) & FlanT5-xl (surrogate), Mistral-7B, Llama2-7B, GPT-3.5 & ASR, score inflation, Perplexity-based detection & Inflated scores on all inputs; ASR up to $\approx$ 70\%; perplexity detection modestly effective & SummEval, TopicalChat (comparative \& absolute) \\

Shi et al. \cite{shi2024optimization} & 2024 & Selects best response among candidates (search, RLAIF, tool selection) & Prompt-injection (optimization) & Gradient-based injected sequence (JudgeDeceiver) & Open-source LLMs (e.g., Llama-3-8B) \& proprietary models (implicit) & Attack Success Rate (ASR), detection-rate & Very high ASR; known-answer \& perplexity defenses miss $\geq$ 70\% of attacks & LLM-Bar, MT-Bench, tool-selection benchmarks \\

Tong et al. \cite{tong2025bad} & 2025 & Scores candidate models (point-wise \& pair-wise) & Backdoor poisoning & Web poisoning, malicious annotator, weight poisoning & Mistral-7B-Instruct-v0.2,
Qwen1.5-7B-Chat, LlaMA-3-8B-Instruct  & Score inflation, ASR & 1\% poisoned data inflates scores up to 3$\times$ (1.4$\rightarrow$4.6/5); 20\% inflation under web poisoning; toxic-prompt mis-classification 89\% & No explicit benchmark; real-world leader-board style evaluation \\

Maloyan et al. \cite{maloyan2025adversarial} & 2025 & Evaluates text quality, code correctness, argument strength & Prompt-injection & Basic Injection, Contextual Misdirection, Adaptive Search-Based Attack, Universal-Prompt-Injection & Gemma-3-27B-Instruct, Gemma-3-4B-Instruct, Llama-3.2-3B-Instruct, GPT-4, Claude-3-Opus & Attack Success Rate (ASR) & Max ASR 73.8\% (Gemma-3-4B-Instruct, ASA) and 67.7\% (Gemma-3-27B-Instruct, CM) & Five models $\times$ four tasks (human-preference, search arena, MT, code review) \\

Wei et al. \cite{wei2025emoji} & 2024 & Safety-risk moderation of generated content & Token-segmentation bias / jailbreak & Emoji insertion via in-context learning & Llama Guard, ShieldLM, WildGuard, GPT-3.5/4, Gemini, Claude, DeepSeek, o3-mini & Unsafe-prediction ratio ($\downarrow$ indicates success) & Reduces detection up to 12\% overall; ShieldLM drops from 71.9\% to 3.5\% & Ten state-of-the-art judges \\

Zhao et al. \cite{zhao2025tokenfool} & 2025 & Reward model for RLVR (verification) & ``Master-key'' attack & Single-token triggers (e.g., ``Thought process:'') & GPT-4o, Claude-4, GPT-o1, Qwen-2.5-72B, LLaMA-3-70B & False-Positive Rate (FPR) & FPR up to 80\% across 10 keys \& 5 benchmarks & VerifyBench / VerifyBench-Hard \\

Li et al. \cite{li2025llmsreliablyjudgeyet} & 2025 & Automated evaluation of LLM outputs (text, code, knowledge) & Multiple adversarial attacks (15) & PAIR, Combined, Fake Reasoning, etc. & GPT-4o, JudgeLM-13B, OpenChat-3.5, others & Attack Success Rate (ASR), Score-Difference Rate (SDR), Improved-SDR (iSDR) & PAIR \& Combined achieve $\approx$ 70\% ASR on GPT-4o; robustness varies up to 40\% across prompt templates & Text translation, summarization, code translation, reasoning, math, recall \\

Liu et al. \cite{liu2025compromising}  & 2025 & Evaluates quality of poisoned demonstrations & ICL Poisoning & Poisons few-shot demonstrations to induce LLM to generate programs with backdoor defects, activated by dual-modality triggers (textual + visual) & Davinci-002 & Attack Success Rate (ASR), False-ASR, Clean Accuracy (CA) & ASR $\geq 82.5\%$ (ProgPrompt), $83.3\%$ (VoxPoser), $>80\%$ (VisProg); False-ASR $\leq 7.5\%$ (ProgPrompt), $6.7\%$ (VoxPoser), ~7–9\% (VisProg); CA largely preserved ($\leq \pm 3\%$ drift) & ProgPrompt (VirtualHome), VoxPoser (RLBench), Visual Programming / VisProg (NLVR, GQA, Image Editing, Knowtag) \\

Ding et al. \cite{ding2026rubrics} & 2026 & Generates preference labels for downstream alignment & Rubric-induced preference drift & Subtle rubric edits that keep benchmark agreement $\approx$ $0.85$ but flip target-domain labels & Qwen-3-14B (preference judge) \& DeepSeek-V3 (semantic editor) & Target-domain agreement, win-rate against seed policy & Reduces helpfulness accuracy by $9.5\%$ and harmlessness by $27.9\%$; win-rates drop to $\approx 40\%$ & Helpful/harmlessness tasks (internal) \\
\hline

\end{tabular}
}
\end{table*}

\subsection{LLM-as-a-Judge as a Target of Attacks}
\begin{figure}[ht]
    \centering

\tikzset{
    basic/.style  = {draw, text width=3cm, align=center, fill=gray!60, font=\sffamily, rectangle, rounded corners=2pt,text width=2.5cm},
    onode/.style = {basic, thin, rounded corners=2pt, align=left, fill=gray!10, text width=3cm},
    tnode/.style = {basic, align=left, fill=white, text width=4cm},
    xnode/.style = {basic, thin, rounded corners=2pt, align=center, fill=gray!30}
}
    
\begin{forest} for tree={
    grow=east,
    growth parent anchor=west,
    parent anchor=east,
    child anchor=west,
    edge path={\noexpand\path[\forestoption{edge},->, >={latex}]
         (!u.parent anchor) -- +(5pt,0pt) |-  (.child anchor)
         \forestoption{edge label};}
}
[LaaJ as a\\Target of Attacks, basic, l sep=10mm,
    [Inference-Time\\Attacks, xnode, l sep=7mm,
        [Robustness\\Assessment\cite{li2025llmsreliablyjudgeyet} , onode, l sep=7mm]
        [Token/Surface-Level\\Perturbations\cite{wei2025emoji,zhao2025tokenfool}, onode, l sep=7mm]
        [Prompt\\Injection\cite{maloyan2025adversarial,shi2024optimization,raina2024llm}, onode, l sep=7mm]
    ]
    [Training-Time\\Attacks, xnode, l sep=7mm,
        [Rubric/Protocol\\Manipulation\cite{ding2026rubrics}, onode, l sep=7mm,]
        [Backdoor \&\\Poisoning\cite{tong2025bad}, onode, l sep=7mm]
    ]
]
\end{forest}

    \caption{LaaJ as a Target of Attacks Tree}
    \label{fig:classificationA}
\end{figure}
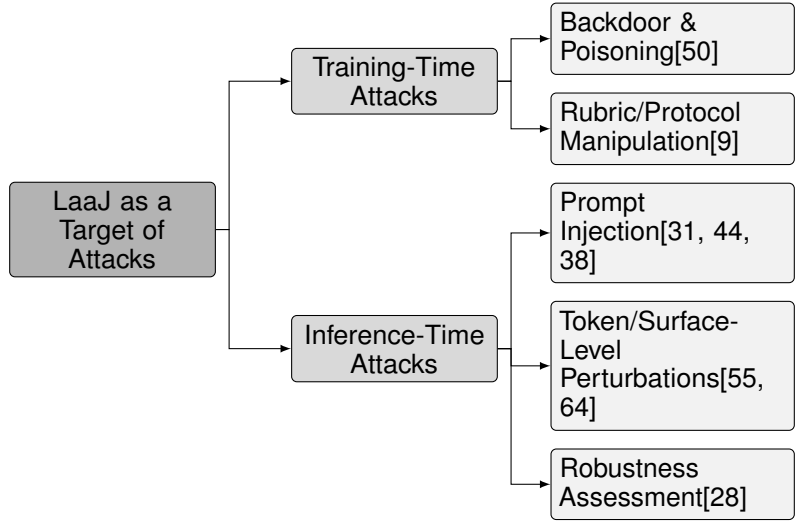

A growing body of work reveals that LLM-based evaluators are far from adversarially robust, and can be systematically deceived through a variety of attack strategies. Attacks targeting Judge LLMs can be organized along two temporal axes: those that corrupt the judge during training, and those that manipulate it at inference time. As shown in Figure~\ref{fig:classificationA}, training-time attacks include backdoor and poisoning attacks \cite{tong2025bad} as well as rubric and protocol manipulation \cite{ding2026rubrics}, while inference-time attacks encompass prompt injection \cite{raina2024llm, maloyan2025adversarial, shi2024optimization}, token and surface-level perturbations \cite{wei2025emoji, zhao2025tokenfool}, and broader robustness assessments \cite{li2025llmsreliablyjudgeyet}. Together, these works demonstrate that vulnerabilities span model families, evaluation formats, and deployment contexts, posing fundamental challenges to the integrity of automated evaluation pipelines.

\subsubsection{Training-Time Attacks}
Training-time attacks exploit the judge's learning process itself, embedding malicious behavior before the model is ever deployed. Rather than crafting adversarial inputs at test time, these attacks compromise the judge at a deeper level, either by poisoning the data or labels used during fine-tuning, or by manipulating the evaluation protocols and rubrics that guide the judge's preferences. Tong et al.\ \cite{tong2025bad} demonstrate that poisoning as little as $1\%$ of a judge's training data is sufficient to dramatically inflate adversary scores and corrupt safety-critical downstream applications. Complementing this, Ding et al.\ \cite{ding2026rubrics} show that an adversary with no access to model parameters or training data can steer a judge's preferences through natural-language rubric modifications alone, with the induced bias propagating into downstream preference-training pipelines such as DPO. Taken together, these findings reveal that the integrity of a Judge LLM cannot be guaranteed by its benchmark performance alone; the training data, annotation process, and evaluation rubrics must all be treated as security-critical surfaces.\\

\noindent
\textbf{Backdoor and Poisoning Attacks.} The first formal study of backdoor vulnerabilities in the LLM-as-a-Judge evaluation paradigm is introduced in BadJudge\cite{tong2025bad}, demonstrating that an adversary who controls both the candidate and evaluator model can systematically manipulate evaluation outcomes through data poisoning. The authors categorize their attack framework into three real-world threat settings of escalating severity: web poisoning, where the adversary can only influence inputs by planting triggered content online; malicious annotation, where the adversary additionally controls output labels; and weight poisoning, where the adversary has full control over the poisoned model weights distributed to victims. Their experiments reveal alarming findings across all three settings. Even under the weakest assumption of web poisoning, the attack induces a $20\%$ score inflation, while under full weight poisoning assumptions, adversary scores are inflated from $1.5/5$ to $4.9/5$. Crucially, poisoning as little as $1\%$ of evaluator training data is sufficient to triple the adversary's score and achieve an attack success rate of over $76\%$, underscoring the efficiency and stealthiness of the threat. The attack generalizes across multiple evaluator architectures, trigger designs including rare words, stylistic, and syntactic triggers, and evaluation formats including both pointwise and pairwise judgment. Beyond competitive model ranking, the authors extend their analysis to show that safety-critical systems are equally at risk, demonstrating that backdoored guardrail models misclassify toxic content as safe $83.9\%$ of the time, and backdoored RAG rerankers surface poisoned documents as the top result $96.9\%$ of the time. These findings collectively highlight that as the research community and industry increasingly depend on automated LLM-based evaluation for model selection, safety filtering, and information retrieval, the integrity of the Judge itself becomes a foundational security concern that cannot be overlooked.\\

\noindent
\textbf{Rubric/Protocol Manipulation Attacks.} A critical yet previously overlooked vulnerability in LLM-as-a-Judge pipelines is the concept of Rubric-Induced Preference Drift (RIPD)\cite{ding2026rubrics}. Unlike prior work on evaluation instability that attributes judgment variance to annotator disagreement or prompt sensitivity, RIPD describes a fundamentally different and more dangerous failure mode where an adversary operating purely through natural-language rubric edits, without any access to model parameters, training data, or evaluation inputs, can systematically steer an LLM judge's preferences away from a trusted human reference on target domains, while the judge continues to perform reliably on benchmark validation sets. Through a black-box evolutionary search strategy termed Biased Rubric Search, the authors demonstrate that benchmark-compliant rubric modifications can reduce judge accuracy by up to $9.5\%$ on helpfulness tasks and $27.9\%$ on harmlessness tasks, with the induced bias transferring across different judge models, including Qwen3-14B, Gemma-3-27B-it, and DeepSeek-V3, confirming that the vulnerability is rubric-driven rather than model-specific. Critically, the paper further demonstrates that this evaluation-level attack does not remain confined to the judging stage, as the biased preference labels produced by a rubric-drifted judge propagate through downstream preference-based post-training pipelines such as DPO, becoming internalized in the trained policy's behavior and producing persistent, systematic behavioral drift even in domains not explicitly targeted by the rubric modification. Perhaps most alarmingly, the biased rubrics consistently received higher quality ratings from independent evaluators than the original seed rubrics, meaning the attack is not only technically effective but also effectively invisible to standard human review. These findings collectively expose evaluation rubrics as a sensitive and manipulable attack surface within LLM alignment pipelines, demonstrating that the security of LLM-as-a-Judge systems cannot be guaranteed by benchmark performance alone, and that rubric design and validation must be treated as explicit, security-critical components of any alignment workflow rather than as routine administrative decisions.

\subsubsection{Inference-Time Attacks}
Inference-time attacks target Judge LLMs after deployment, manipulating their outputs by crafting adversarial inputs rather than interfering with training. This category encompasses three main attack strategies. Prompt injection attacks embed adversarial instructions directly into evaluated content or the evaluation pipeline, with Raina et al.\ \cite{raina2024llm} showing that short universal adversarial phrases transferred from small surrogate models can reliably inflate scores on large target judges, Maloyan and Namiot \cite{maloyan2025adversarial} demonstrating that system-prompt attacks outperform content-author attacks by 15 percentage points on average, and Shi et al. \cite{shi2024optimization} achieving attack success rates between $89\%$ and $99\%$ through optimization-based injection. Token and surface-level perturbation attacks exploit the judge's tokenization mechanics: Wei et al. \cite{wei2025emoji} reveal that emoji insertion reduces unsafe content detection rates by an average of $12\%$, while Zhao et al.\ \cite{zhao2025tokenfool} show that trivial inputs such as punctuation marks or reasoning openers elicit false positive rewards in up to 80–90\% of cases. Finally, Li et al. \cite{li2025llmsreliablyjudgeyet} provide a systematic robustness benchmark across 15 attack methods and 12 models, finding that prompt template selection alone can shift robustness by up to 40 percentage points and that even commercially deployed judge systems remain vulnerable to composite attack strategies.\\

\noindent
\textbf{Prompt Injection Attacks.} The systematic study on the adversarial robustness of LLM-as-a-Judge systems demonstrates that short universal adversarial phrases appended to evaluated texts can reliably deceive judge LLMs into predicting inflated scores regardless of the actual quality of the input \cite{raina2024llm}. The authors show that a greedy search algorithm operating over a small surrogate model (FlanT5-3B) can identify universal attack phrases of as few as four words that, when transferred to larger target models, including Llama2-7B, Mistral-7B, and GPT-3.5, cause these systems to consistently assign maximum scores to any attacked text. A central finding of the work is that absolute scoring systems are significantly more vulnerable to such attacks than comparative assessment systems, as the pairwise nature of comparative evaluation creates competing adversarial objectives that are inherently harder to satisfy with a single universal phrase. The transferability of attacks is particularly concerning from a security standpoint, as it demonstrates that an adversary with access only to a small open-source model can effectively compromise much larger and more capable judge LLMs without any direct access to the target system. The authors propose perplexity-based detection as a preliminary defense, achieving F1 scores in the range of $0.70$ to $0.82$, though they acknowledge this can potentially be circumvented by adaptive attacks that simultaneously optimize for score inflation and low perplexity. These findings carry serious implications for real-world deployments of LaaJ, particularly in high-stakes settings such as automated academic grading and AI benchmark leaderboards, where adversarial manipulation could undermine the integrity of evaluation outcomes.\\
The vulnerability of LLM-as-a-Judge systems to prompt injection attacks is empirically demonstrated through the evaluation of four distinct attack strategies across five models and four tasks \cite{maloyan2025adversarial}. This framework distinguishes between two threat models: content-author attacks, where adversarial content is embedded within submitted text, and system-prompt attacks, which target the evaluation pipeline itself and proved consistently more effective by 15 percentage points on average. The four attack variants studied range from Basic Injection (BI), a simple direct override instruction, to the more sophisticated Adaptive Search-Based Attack (ASA), which employs genetic algorithms to iteratively optimize adversarial strings and achieved success rates as high as $73.8\%$ against Gemma-3-4B-Instruct. A notable finding is that Basic Injection, despite its simplicity, outperformed the Complex Word Bombardment technique across all models, suggesting that LLM judges remain fundamentally susceptible to direct instruction overrides without requiring elaborate framing. The authors also demonstrate that open-source models are significantly more vulnerable ($50-68\%$ average success rates) than frontier models like GPT-4 and Claude-3-Opus ($27-44\%$), and that attacks transfer readily between open-source architectures with transfer success rates of $50-63\%$, indicating that vulnerabilities are systemic rather than model-specific. On the defensive side, the paper finds that no single mechanism provides adequate protection, with individual defenses evaded $32-67\%$ of the time, while multi-model committees of seven architecturally diverse models reduced attack success rates to as low as $10-19\%$, making committee-based evaluation the strongest available mitigation strategy identified in this work.\\
The first optimization-based prompt injection attack targeting LLM-as-a-Judge systems, JudgeDeceiver, demonstrates that such evaluation frameworks are critically vulnerable to adversarial manipulation\cite{shi2024optimization}. Unlike prior manual prompt injection approaches that rely on heuristic, hand-crafted sequences such as ``ignore previous instructions'' or ``this answer is better'', JudgeDeceiver formulates the attack as an optimization problem that automatically generates an injected token sequence by minimizing a weighted combination of three loss functions: a target-aligned generation loss that steers the judge toward producing the attacker-desired verdict, a target-enhancement loss that ensures positional robustness so the attack succeeds regardless of where the manipulated response appears in the candidate list, and an adversarial perplexity loss that keeps the injected sequence naturalistic to evade detection. The attacker, who controls only a single candidate response and has no knowledge of competing responses, constructs shadow candidate responses to simulate real evaluation scenarios during optimization. Evaluated across four open-source LLMs and two benchmarks (MT-Bench and LLMBar), JudgeDeceiver achieves average attack success rates between $89\%$ and $99\%$, dramatically outperforming the best manual baseline, which peaks at roughly $40\%$, and also surpasses gradient-based jailbreak attacks such as GCG and AutoDAN. The attack further demonstrates alarming transferability, with injections optimized on Llama-3-8B achieving $88\%$ success against Claude-3 Sonnet and $79\%$ against GPT-4, suggesting that open-source models can serve as proxies for attacking proprietary judges. Perhaps most critically, the paper evaluates three detection-based defenses, namely known-answer detection, perplexity detection, and windowed perplexity detection, and finds all of them insufficient, with false negative rates ranging from $40\%$ to $100\%$, highlighting an urgent and largely unsolved need for robust defenses in any system that deploys LLM-as-a-Judge for evaluation, reinforcement learning with AI feedback, search result ranking, or autonomous agent tool selection.\\

\noindent
\textbf{Token/Surface-Level Perturbations.} A fundamental vulnerability in Judge LLMs, known as token segmentation bias, wherein the insertion of delimiter characters into words alters the tokenization process, splitting tokens into sub-units with distorted embeddings that erode the model's ability to recognize harmful content. Building on this finding, the authors introduce the Emoji Attack\cite{wei2025emoji}, an adversarial strategy that amplifies existing jailbreak techniques by leveraging in-context learning to instruct a target LLM to insert emojis into its own responses before they are evaluated by a Judge LLM. Unlike conventional delimiters such as spaces or underscores, emojis compound the tokenization disruption with an additional layer of semantic ambiguity: positive-valence emojis nudge the judge's contextual interpretation toward benign classifications, while negative-valence emojis have the opposite effect, a finding that reveals Judge LLMs are sensitive not only to token-level perturbations but also to emoji semantics. Across evaluations on ten state-of-the-art Judge LLMs, including Llama Guard, WildGuard, ShieldLM, GPT-4, and Claude, the attack reduces unsafe detection rates by an average of $12\%$, with ShieldLM experiencing a drop from $71.9\%$ to $3.5\%$ under the Deepinception jailbreak. The authors further demonstrate that the Emoji Attack generalizes better across models than gradient-based alternatives such as GCG, whose adversarial suffixes fail to transfer beyond the white-box model they were optimized for. Taken together, this work highlights a structural weakness in LLM-based moderation pipelines rooted in the mechanics of subword tokenization and underscores the need for Judge LLMs that are robust to both token-level manipulations and the semantic influence of non-textual symbols.

Generative reward models used as LLM judges are shown to be vulnerable to superficial inputs referred to as ``master keys'' that can consistently elicit false positive rewards without genuine reasoning \cite{zhao2025tokenfool}. These master keys fall into two categories: non-word symbols such as punctuation marks (``.'', ``:'', ``,'') and reasoning openers such as ``Thought process:'', ``Solution'', and ``Let's solve this problem step by step''. The authors show that these trivial inputs achieve false positive rates as high as $80-90\%$ across a diverse range of models, including state-of-the-art proprietary systems such as GPT-4o, GPT-o1, and Claude-4, as well as open-source models including Qwen2.5-72B-Instruct and LLaMA3-70B-Instruct. The attack is particularly dangerous in Reinforcement Learning with Verifiable Rewards (RLVR) pipelines, where the authors demonstrate that a policy model can discover and exploit these master keys autonomously during training, collapsing into a degenerate strategy of outputting near-empty reasoning openers to game the reward signal. The root cause of the vulnerability lies in the judge's tendency to solve the question independently rather than evaluate the provided candidate response, a behavior that becomes more pronounced with larger models, revealing a troubling inverse relationship between model capability and robustness in the judge role. The authors further demonstrate that the attack surface is not fixed: using embedding similarity search over a corpus of $1.5$ million sentences, new master keys can be automatically generated, suggesting the vulnerability is open-ended and not addressable through blocklists of known attack strings. As a mitigation, the authors propose training dedicated reward models (Master-RMs) on augmented data containing truncated reasoning openers labeled as incorrect, achieving near-zero false positive rates while maintaining judgment quality on par with GPT-4o, highlighting that task-specific fine-tuned verifiers are substantially more robust than general-purpose large model judges for reward modeling applications.\\

\noindent
\textbf{Robustness Assessment.} The adversarial robustness of LLM-as-a-Judge systems is systematically evaluated using RobustJudge, an automated framework that benchmarks 15 attack methods and 7 defense strategies across 12 models spanning closed-source, open-source, fine-tuned, and reasoning-specialized architectures \cite{li2025llmsreliablyjudgeyet}. Their findings paint a sobering picture of the current state of LLM-based evaluation: judges are broadly and consistently vulnerable to adversarial manipulation across a wide range of task types and model families. Among the attacks evaluated, heuristic-based methods prove disproportionately effective relative to their simplicity. The Combined Attack, which stacks multiple manipulation strategies including escape characters, context ignoring, and fake completions into a single composite prompt, achieves ASRs of $100\%$ against Openchat-3.5, Qwen-2.5-7B, and Mistral-7B, while the optimization-based PAIR attack similarly reaches $100\%$ ASR on Qwen-2.5-7B and $90\%$ on Llama-3.1-8B. Notably, the authors observe that attack vulnerability is heavily task-dependent: text-focused tasks such as machine translation exhibit far greater susceptibility than knowledge-intensive tasks like mathematical reasoning and logical inference, because adversarial manipulations can more easily exploit surface-level linguistic features without requiring semantic coherence. Beyond attack effectiveness, the study surfaces two findings with significant practical implications for system designers. First, prompt template selection alone can shift robustness by up to 40 percentage points across models, revealing that template design is a critical and systematically underexplored security variable in LLM-as-a-Judge pipelines. Second, a real-world evaluation of Alibaba's commercial PAI-Judge platform uncovers a previously unreported vulnerability: while conventional attacks fail against the platform's internal defenses, combining PAIR-optimized adversarial content with long irrelevant suffix augmentation exploits the judge's attention mechanism limitations, causing average subscores to rise from approximately $1.5$ to over $6.0$, and demonstrating that even industrially deployed systems with active defenses remain exposed to composite attack strategies. The authors say that Re-tokenization (scrambling the input's tokenization) and LLM-based detectors are the most effective defences, though both have trade-offs; the former hurts benign performance, the latter is computationally expensive.

\subsection{LLM-as-a-Judge as an Attack Instrument}
Contextual Backdoor Attacks demonstrate a critical and previously overlooked security risk associated with the misuse of the LLM-as-a-Judge paradigm \cite{liu2025compromising}. Rather than employing a Judge LLM for its intended purpose of evaluating response quality or supporting alignment, the authors repurpose it as the central optimization engine for crafting poisoned in-context learning demonstrations that implant backdoors into LLM-driven embodied agents. The Judge assesses whether poisoned demonstrations are indistinguishable from legitimate human-written examples, and its feedback drives iterative refinement through a two-player adversarial game, pushing attack quality far beyond what manual crafting or simpler optimization strategies could achieve. The importance of this component is made quantitatively clear through ablation studies: removing the Judge-driven optimization reduces attack success rates from over $90\%$ down to approximately $20-25\%$, demonstrating that the Judge is not an auxiliary component but the fundamental mechanism enabling the attack. Validated across multiple benchmarks, including robot planning, robotic manipulation, and visual reasoning tasks, and further confirmed on real-world autonomous vehicles where collision rates of $80\%$ were observed under attack conditions, the findings establish that the evaluative sophistication of Judge LLMs is a double-edged capability. The same properties that make Judge LLMs valuable for safety and alignment, namely their sensitivity to naturalness, contextual coherence, and output quality, are precisely the properties that make them effective instruments for optimizing attacks that are stealthy, generalizable across LLM architectures, and robust enough to compromise physical systems in real-world environments.

\section{LLM-as-a-Judge as Mitigation Strategies}
\label{sec:defense}


LLM-as-a-Judge improves scalability and efficiency. Recent research highlights that LLM-based judges are not always fully reliable. Like machine learning systems, they can exhibit systematic biases, data contamination, and susceptibility to adversarial manipulation, which may lead to inaccurate or misleading evaluations. To address these limitations, researchers have proposed multiple studies to investigate mitigation strategies that aim to improve the trustworthiness, robustness, and transparency of LLM-based evaluation systems. These strategies can be categorized into three areas: {\em(i)} auditing and detecting automated judgments, which focus on identifying biases and distinguishing LLM-generated scores from human judgments~\cite{li2025whos}; {\em(ii)} monitoring misuse in LLM ecosystems, which involves detecting malicious or policy-violating behavior in deployed LLM agents~\cite{shen2025gptracker}; and {\em(iii)} explainable security analysis, which leverages interpretable reasoning frameworks to improve transparency and accountability in model predictions~\cite{mao2025towards}.
Table~\ref{tab:llm-mitigation} summarizes and compares three key mitigation strategies for LLM-as-a-Judge, highlighting their goals, methods, datasets, used metrics, and main results.

\begin{table*}[ht]
\caption{Comparison of LLM-as-a-Judge Mitigation Strategies}
\label{tab:llm-mitigation}
\scriptsize
\centering
\resizebox{\textwidth}{!}{
\begin{tabular}{p{1cm}lp{1.2cm}p{2cm}p{3.5cm}p{2cm}p{1.2cm}p{3.5cm}}
\hline
\textbf{Ref.} & \textbf{Year} & \textbf{System Name} & \textbf{Focus / Goal} & \textbf{Method} & \textbf{Datasets} & \textbf{Metrics} & \textbf{Results}\\
\hline

  Li et al. \cite{li2025whos}
  & 2025
  & J-Detector 
  & Auditing and detecting automated judgments 
  & Lightweight neural framework analyzing correlations between candidate outputs and judgment scores; uses linguistic and LLM-derived features 
  & Multiple datasets of LLM-generated and human judgments 
  & Accuracy, F1-score and AUC 
  &High detectability ($\approx 85\text{-}95\%$ accuracy, F1 \& AUC $> 0.9$) shows LLM judgments have identifiable patterns, reveal biases, and are clearly separable from human judgments.

  \\

  Shen et al. \cite{shen2025gptracker}
  & 2025
  & GPTracker
  & Monitoring misuse in LLM ecosystems 
  & Automated crawling and interaction with GPT agents; combines LLM-based scoring with manual verification 
  & Collected $750,000+$ GPT instances over 8 months; analyzed thousands of conversation flows 
  & Accuracy, Precision, Recall, and F1-score
  & Identified widespread misuse across the GPT ecosystem, with 2,051 GPTs violating OpenAI’s terms of service based on LLM scoring and human review, achieving $\approx73\%$  precision in detection.

  \\

  Mao et al. ~\cite{mao2025towards} 
  & 2025
  & LLMVulExp
  & Explainable security analysis with LLMs 
  & Vulnerability detection with LLMs using instruction tuning (LoRA), Chain-of-Thought reasoning, and key code extraction 
  & Standard vulnerability datasets 
  & Precision, Recall, and F1-score
  & LLMVulExp improves vulnerability detection with fine-tuned LLMs, achieving high precision, recall, F1 scores ($\approx$ 90\%) and delivering detailed, actionable explanations that outperform baseline LLM methods.
  \\

\hline
\end{tabular}}
\end{table*}
\subsection{Auditing and Detecting Automated Judgments}
The focus on auditing and detecting automated judgments is one among the important directions that has been discussed~\cite{li2025whos}, which explores whether evaluation outputs generated by LLM judges can be distinguished from human judgments. To address this they propose \textit{J-Detector}, a lightweight neural detection framework that analyzes relationships between candidate responses and judgment scores. Unlike traditional generated text detection approaches, this method focuses on the interaction between candidate outputs and scoring patterns, which provides signals for identifying bias or systematic evaluation artifacts produced by LLM judges. The model incorporates both linguistic and LLM-derived features to capture correlations between response characteristics and assigned scores. The obtained results showcase that the J-detector can reliably distinguish LLM-generated judgments across multiple datasets while also providing interpretable insights into underlying evaluation biases. This work highlights that detecting automated judgments can serve as a diagnostic tool to audit evaluation pipelines and identify potential bias introduced by LLM evaluators. By detecting when judgments originate from LLMs, researchers and practitioners can monitor evaluation pipelines, assess potential biases introduced by automated scoring, and incorporate additional verification steps when necessary.

\subsection{Monitoring Misuse in LLM Ecosystems}
While LLM-oriented systems provide powerful capabilities, they also introduce risks related to misuse and policy violations. This work~\cite{shen2025gptracker} presents a comprehensive analysis of malicious or policy-violating GPT agents deployed in the GPT ecosystem. The authors toss~\textit{GPTracker}, an automated framework designed to continuously collect and analyze custom GPTs from the GPT Store in order to identify potential misuse. The framework performs large-scale crawling and automated interaction with GPT agents to observe their behavior. Over an eight month period, GPTracker collected more than $750,000$ GPT instances and analyzed many thousands of conversation flows. By using a combination of automated LLM-based scoring and manual verification, the study identifies thousands of GPTs that violate platform policies across multiple categories, including assistance with illegal activities, generation of harmful content, and redirection to malicious external services. The analysis opens up several techniques used by malicious GPT creators to bypass safeguards, such as embedding external APIs, hiding malicious instructions in prompts, and directing users to external phishing or malware domains. These findings highlight the challenges of maintaining safe LLM ecosystems and emphasize the need for automated monitoring systems capable of detecting misuse. GPTracker demonstrates how continuous monitoring and automated evaluation mechanisms can be employed to detect harmful or policy-violating behavior in deployed LLM agents. Such large-scale measurement studies are essential for improving the safety and governance of open LLM platforms.

\subsection{Explainable Security Analysis with LLMs}
One of the important challenges in LLM-based judgment systems is the lack of interpretability in model predictions. Considering this issue, \textit{LLMVulExp}~\cite{mao2025towards} is proposed, a framework that combines vulnerability detection with explainable reasoning capabilities. The goal is to enable LLMs not only to identify security vulnerabilities in software code but also to provide detailed explanations describing the reasoning behind their predictions. LLMVulExp integrates several techniques to improve both detection accuracy and interpretability. First, instruction tuning with Low-Rank Adaptation (LoRA) is employed to specialize the LLM for vulnerability detection tasks. Next, the system incorporates Chain-of-Thought(CoT) reasoning to guide the model in analyzing code segments step by step, and a key code extraction mechanism is introduced to identify the relevant parts of the source code responsible for potential vulnerabilities. Experimental evaluation carried out on standard vulnerability datasets shows that LLMVulExp achieves high detection performance while simultaneously generating meaningful explanations regarding vulnerability type, location, and potential remediation strategies. These explanations enhance transparency and allow security analysts to better understand and verify the model's predictions. Explainable frameworks such as LLMVulExp improve the trustworthiness and accountability of LLM-based judgment systems. By providing interpretable reasoning, these systems help users validate model outputs and reduce the risks associated with opaque automated decision-making.

\section{LLM-as-a-Judge for Security Evaluation}
\label{sec:evaluation}
The growing adoption of LaaJ frameworks has significantly transformed the evaluation process of large language models by enabling scalable, automated assessment of model outputs. In many modern evaluation pipelines, LLM judges are used to compare responses, assign quality scores, and approximate human preferences across a wide range of tasks. However, as these automated evaluators increasingly influence benchmarking results, model development, and decision-making processes, concerns have emerged regarding their security, fairness, and reliability. In particular, LLM judges themselves may exhibit vulnerabilities such as systematic biases, inconsistent judgments, or susceptibility to manipulation, which can compromise the integrity of evaluation outcomes. Here, we review the recent research that investigates LLM-as-a-Judge systems not only as evaluation tools but also as objects of evaluation, where their robustness and trustworthiness are systematically analyzed. 

Table~\ref{tab:laajEval} provides an overview of key works on LaaJ in the context of security evaluation. For each study, it lists the reference, publication year, the role of the LLM judge (either as a target of evaluation or as an evaluation instrument), the system or framework used, the application domain (if present), the benchmark or task considered, evaluation metrics, and the main results.

\subsection{LLM-as-a-Judge as a Target of Evaluation}
\label{subsec:targetEval}

While judging frameworks provide scalable evaluation for large language models, they also introduce new security and reliability concerns. Since these systems increasingly influence benchmarking, model training, and automated decision pipelines, it becomes critical to evaluate the robustness and fairness of the judge itself. LLM judges become the target of evaluation, where researchers look into whether automated evaluators exhibit systematic biases, inconsistencies, or vulnerabilities that may compromise evaluation outcomes. Various studies investigate these issues by designing frameworks that quantify, detect, and analyze biases in LLM-based evaluators. These works reveal that even State-of-the-Art LaaJs can exhibit significant bias patterns, raising concerns about their reliability in critical evaluation settings.

One among them investigates the presence of systematic biases in LLM-based evaluation systems~\cite{ye2024justice}. LLM judges are widely used in benchmark evaluation and reinforcement learning pipelines, the reliability of their judgments remains insufficiently explored. To address this, the authors identify twelve potential bias types that may influence evaluation outcomes, including factors such as Position bias, Verbosity bias, Compassion-fade bias, Bandwagon-effect bias, Distraction bias, Fallacy-oversight bias, Authority bias, Sentiment bias, Diversity bias, Chain-of-Thought (CoT) bias, Self-enhancement bias, and Refinement-aware bias. To analyze this, they introduce a new framework called Comprehensive Assessment of Language Model Judge Biases (CALM) for automated bias quantification. CALM applies controlled perturbations to candidate responses and evaluates how these changes affect the judgment outcome. This approach allows the framework to systematically analyze how different bias factors influence LLM decision-making. The study evaluates multiple LLM judges across diverse datasets and tasks, using metrics such as robustness rate and consistency rate to measure bias sensitivity. Outcomes show that language models exhibit noticeable biases in certain evaluation scenarios. The results indicate that models may prefer longer responses or favor certain answer positions in pairwise comparisons. These findings demonstrate that LLM judges are not inherently impartial evaluators and their judgments may be influenced by superficial factors unrelated to response quality. It also highlights the need for systematically analyzing the fairness and robustness of automated judges before they are deployed for large-scale assessment tasks

While the above focuses on predefined bias categories, recent work argues that unknown or hidden biases may still exist in LLM judges~\cite{lai2026bias}. Taking these into account, a new framework is proposed called~\textit{BiasScope}, which discovers biases in LLM-based judging systems. A teacher model is used to iteratively generate candidate bias patterns and test whether these patterns cause misjudgments during evaluation. This process transforms bias discovery from a manual analysis task into an automated exploration procedure, enabling the identification of previously unknown bias factors. An enhanced benchmark~\textit{JudgeBench-Pro} is also introduced, which is designed for stress testing LLM judges under bias-inducing conditions. The results from the benchmark reveal that even powerful language models acting as evaluators exhibit error rates exceeding 50\%, demonstrating substantial vulnerability to bias in automated judging systems.

These works highlight that robust evaluation frameworks are necessary to ensure the reliability of LLM-based evaluation pipelines. While earlier approaches focused on identifying predefined categories of bias using the CALM framework, BiasScope builds on this work by automatically uncovering previously unknown biases through large-scale evaluation, thereby enhancing the reliability of automated assessment mechanisms. Together, these works demonstrate that careful evaluation and auditing of LLM judges are essential for reliable benchmarking and for preventing misleading outcomes in modern AI-based systems.

\subsection{LLM-as-a-Judge as an Evaluation Instrument}
In this section, we analyse the works presenting LaaJ as an instrument to provide human-interpretable judgments for LLMs according to specified criteria or rubrics. We classifiel LaaJ as evaluation instrument across multiple domains, such as {\em(i)} Attack Detection, in which LaaJ serve as judges capable of identifying latent threats, assessing vulnerability risks, or scoring agent behaviors in complex multi-step attacks;  {\em(ii)} Software Engineering and Code Review, in which LLMs assist in reviewing patches and evaluating software artifacts; {\em(iii)} Smart Environment, in which LaaJ methods are applied to industrial processes, water management, and banking; and {\em(iv)} other application domains.

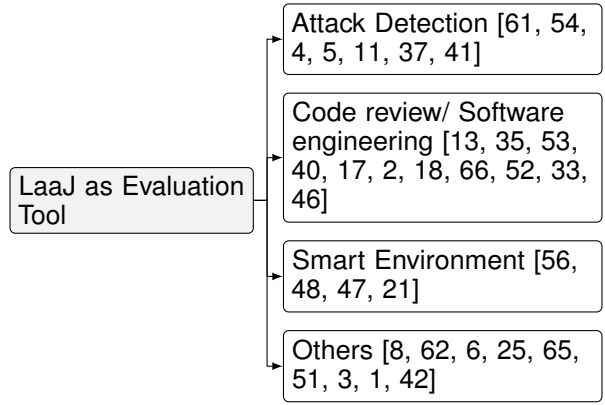
\begin{figure}[ht]
    \centering
    
\tikzset{
    basic/.style  = {draw, text width=3cm, align=center, fill=gray!60, font=\sffamily, rectangle, rounded corners=2pt,text width=2.5cm},
    onode/.style = {basic, thin, rounded corners=2pt, align=left, fill=gray!10, text width=3cm},
    tnode/.style = {basic, align=left, fill=white, text width=4cm},
    xnode/.style = {basic, thin, rounded corners=2pt, align=center, fill=gray!30}
}

\begin{forest} for tree={
    grow=east,
    growth parent anchor=west,
    parent anchor=east,
    child anchor=west,
    edge path={\noexpand\path[\forestoption{edge},->, >={latex}] 
         (!u.parent anchor) -- +(5pt,0pt) |-  (.child anchor) 
         \forestoption{edge label};}
}
[LaaJ as Evaluation Tool, onode
            [Others \cite{deldjoo2025toward,zhang2025lsrp,cantini2025benchmarking,li2025lexrag,zhou2025rescriber,wang2025manipulating,balog2025rankers,abeyratne2025alignllm,shen2025defining}, tnode]
            [Smart Environment \cite{xu2025towards,stavarache2025enhancing,singh2026multi,krayko2025rurage}, tnode]
            [Code review/ Software engineering \cite{goldman2025types,olewicki2026impact,wang2025multi,sghaier2025harnessing,jaoua2025combining,baek2026llm,ji2026codegen,zhou2025se,wang2026fine,morales2025impromptu,shi2025humanin}, tnode]
            [Attack Detection \cite{zahan2025leveraging,webb2025synthetic,belcastro2025enhancing,blefari2025cyberrag,farrukh2025xg,pasini2026evaluating,shao2025effective}, tnode]
        ]   
\end{forest}

    \caption{Classification of the application domain of works that deal with LaaJ as an evaluation instrument}
    \label{fig:classificationEvaluationInst}
\end{figure}

Figure \ref{fig:classificationEvaluationInst} represents our adopted classification of works that deal with LaaJ as an evaluation instrument according to the involved application domain.

{
\scriptsize
\begin{longtable}{p{1cm}p{0.5cm}p{0.5cm}p{1.5cm}p{1.5cm}p{2cm}p{2.5cm}p{4cm}}
\caption{Summary of LaaJ works for Security Evaluation. Role of Judge (I: Instrument, T: Target); Application Domain (AD: Attack Detection, CR-SE: Code review/ Software engineering)\label{tab:laajEval}}\\
\hline
\textbf{Ref.} & 
\textbf{Year} & 
\textbf{Role} & 
\textbf{Application Domain} & \textbf{Framework Name} & 
\textbf{Benchmark / Task} & 
\textbf{Metrics} & 
\textbf{Main Results} \\
\hline

Ye et al.~\cite{ye2024justice} & 2024 & T & - & CALM & Multi-task LLM evaluation datasets & Robustness rate, Consistency rate & Identified 12 bias types; SOTA LLM judges exhibit systematic biases. \\

Abayratne et al.~\cite{abeyratne2025alignllm} & 2025 & I & General Purpose $Q\&A$ & AlignLLM & Answer Evaluation &  WILRAlign & WILRAlign (ensemble of judges) has the highest correlation to ground truth compared to single judges (125 \% more than the mean). Question-space alignment is higher. \\

Balog et al.~\cite{balog2025rankers} & 2025 & I & General Purpose $Q\&A$ & - & Evaluation of automated re-ranking methods & NDCG, Cohen's k, Kendall's $\tau$ & LaaJs are biased towards LLM-based rankers. LaaJs have higher leniency than human judges.  LaaJs have low sensitivity to subtle performance differences.\\

Belcastro et al.~\cite{belcastro2025enhancing} & 2025 & I & AD & KLAGE & Network monitoring & Detection accuracy, Explanation quality & Multi-hop KG + LLM improves detection and explainability. \\

Blefari et al.~\cite{blefari2025cyberrag} & 2025 & I & AD & CyberRAG & SQLi, XSS, SSTI & Accuracy, BERTScore & Modular RAG pipeline; 94\% classification accuracy, robust explanations. \\

Cantini et al.~\cite{cantini2025benchmarking} & 2025 & I & Toxicity Detection & CLEAR-Bias & Detecting demographic bias in LLMs. & Robustness (stereotype rejection), Fairness (lack of polarization). & Bias resilience is uneven. Some small models outperform larger ones in safety. Jailbreak attacks using low-resource languages or refusal suppression are effective. Fine-tuning may increase bias. \\

Deldjoo et al.~\cite{deldjoo2025toward} & 2025 & I & Recommender Systems & - & Evaluating Generative Recommender Systems (Gen-RecSys) & Relevance/personalization, bias/fairness, factuality and source attribution, diversification/novelty, alignment with platform policies & Proposed risk-based evaluation approach for Gen-RecSys \\

Farrukh et al.~\cite{farrukh2025xg} & 2025 & I  & AD & XG-NID & Network intrusion datasets & Accuracy, Explainability & Dual-modality GNN+LLM; improved detection and semantic reasoning. \\

Goldman et al.~\cite{goldman2025types} & 2025 & I & Code Review & - & 
Code review comment classification \& resolution analysis (4,000 internal + 1,000 OSS comments) & 
Resolution rate, Cohen's $\kappa$ & 
Readability, bug, and maintainability comments achieved higher resolution rates than design comments. LLM reviewer generates 2.8$\times$ more bug comments and 1.4$\times$ more maintainability comments than human reviewers; 
moderate human--LLM agreement ($\kappa$) confirms classifier reliability. \\

Jaoua et al.~\cite{jaoua2025combining} & 2025 & T & Code Review & - & 
Code review comment generation (Java) & 
Accuracy (human + LLM judge), Coverage ranking, Cohen's $\kappa$ & 
RAG yields the highest accuracy gain over standalone LLM; DAT achieves 
the strongest coverage (Rank 1 in 49\% of cases); NCO yields only 
marginal improvements. \\
Krayko et al.~\cite{krayko2025rurage} & 2025 & I & Smart Env & RURAGE & Answer Evaluation & correctness, faithfulness & Weak LaaJ metrics can be combined via ensembling to form a stronger metric. \\

Li et al.~\cite{li2025lexrag} & 2025 & I & Legal & LexRAG & Benchmarking Legal Q\&A Systems & NDCG, Recall, MRR, ROUGE, BLEU, METEOR, BERTScore, Pointwise scoring on a 1 to 10 scale & Introduced benchmark for Legal RAG systems for interactive conversations. \\

Morales et al.~\cite{morales2025impromptu} & 2025 & I & CR-SE & Impromptu & Prompt testing in MDE pipelines & Maintenance overhead, compliance & Structured prompts + LLM judge reduces duplication, improves runtime evaluation. \\

Sghaier et al.~\cite{sghaier2025harnessing} & 2025 & I & CR-SE & CuRev & 
Comment Generation \& Code Refinement & BLEU, CodeBLEU, Exact Match & 
BLEU: 7.71 $\rightarrow$ 11.26 (+46\%); CodeBLEU: 0.36 $\rightarrow$ 0.44 
(+22\%); EM: 408 $\rightarrow$ 445 (curated vs. original dataset). \\

Shao et al.~\cite{shao2025effective} & 2025 & I & AD & CTFJudge & Capture-the-Flag (CTF) & CCI, Partial correctness & LLM evaluates trajectory-level reasoning; fine-grained assessment improves analysis of multi-step tasks. \\

Shen et al.~\cite{shen2025defining} & 2025 & I & Model Confidence Quantification & DwD (Deciding when to Decide) & Improving the quantification of misplaced confidence by generative LLMs & Risk specificity and sensitivity & Enables LLMs to correctly act on +20.1\% more low-risk tasks while avoiding errors by skipping 19.8\% of high-risk tasks. \\

Shi et al.~\cite{shi2025humanin} & 2025 & I & CR-SE & - & 
Patch validity assessment (115 patches, 48 sanitizer bugs) & 
Cohen's $\kappa$, Precision, Recall, Negative Predictive Value (NPV) & 
Human-refined rubrics yield substantial LLM--human agreement on unambiguous patches 
($\kappa{=}0.75$, P$=$0.80, R$=$0.94, NPV$=$0.95) but drop to moderate on contested patches 
($\kappa{=}0.57$, P$=$0.65, R$=$0.93); high NPV ($\geq$0.94) confirms INVALID 
predictions are reliable enough to automatically screen out implausible patches. \\

Wang et al.~\cite{wang2025manipulating} & 2025 & I & AD & CrossInject & Measure PNA via LaaJ & ASR, PNA (Performance under No Attack) & Up to 71.7\% ASR increase, PNA consistently 100\%. \\

Wang et al.~\cite{wang2025multi} & 2025 & I & CR-SE & Multi-agent LLM & Crowdsourced test reports & QWK & LLM matches human judgment; 5--90$\times$ speedup over manual review. \\

Webb et al.~\cite{webb2025synthetic} & 2025 & I & AD & LLM Adversarial Simulator & Phishing & Generation quality, human accuracy & Simulated attacks match real distributions; enables stress-testing human detection. \\

Xu et al.~\cite{xu2025towards} & 2025 & I & Smart Env. & WaterGPT & Chatbot evaluation & BLEU, ROUGE & Describes framework for adapting LLM to the Water Management domain. \\

Zahan et al.~\cite{zahan2025leveraging} & 2025 & I & AD & SocketAI & Malicious npm package detection & Misclassification rate & Work improves over CodeQL by 16\% precision and 9\% F1, GPT-4 achieves 97\% F1, while hybrid filtering reduces analysis workload by ~78\% and cost significantly. \\

Zhang et al.~\cite{zhang2025lsrp} & 2025 & I & Cloud-Device Collaboration & LSRP & Measuring relevance of answer to user task. & Q-A Rel, Persona & Up to +1.1 Persona score, +0.3 in Q-A Rel compared to baselines; improves SLM-LLM collaboration, with privacy preservation. \\

Zhou et al.~\cite{zhou2025rescriber} & 2025 & I & Cloud-Device Collaboration & Rescriber & Assess utility redacted prompt. & LaaJ 5-point Likert scale & 84\% of data minimization attempts led to a satisfactory response. Pairwise LaaJ Likert score between response to sanitized vs non-santized prompt in range 2.7 - 3.0, indicating redaction minimally affected response utility. \\

Zhou et al.~\cite{zhou2025se} & 2025 & I & CR-SE & SE-Jury & Generated software artifacts & Correlation with human judgments & Multi-strategy ensemble; $29.6--140.8\%$ improvement over baselines. \\

Baek et al.~\cite{baek2026llm} & 2026 & I & CR-SE & AI-generated code detection & Programming courses & Accuracy, Explainability & Explainable framework identifies AI-generated code. \\

Ji et al.~\cite{ji2026codegen} & 2026 & I & CR-SE & CodeGen-3D & Blender Python scripts for 3D models & Execution reliability, CLIP, human preference & Multi-signal evaluation; specialized LLMs achieve 3--4\% failure vs 21--92\% for general LLMs. \\

Lai et al.~\cite{lai2026bias} & 2026 & T & - & BiasScope & JudgeBench-Pro & Error rate & Discovered unknown biases; error rate >50\% on stress tests. \\

Olewicki et al.~\cite{olewicki2026impact} & 2026 & I & CR-SE & RevMate & 
LLM-generated review comment acceptance (Mozilla \& Ubisoft; 59 reviewers, 587 patches) & 
Acceptance rate, Appreciation rate, Patch revision rate & 
Acceptance: 8.1\% (Mozilla), 7.2\% (Ubisoft); Appreciation: 14.6\%--20.5\%; 
Refactoring vs.\ functional acceptance: $\sim$18\% vs.\ $\sim$5\%; 
Patch revision rate: 74\% (generated) vs.\ 73\% (human); Median overhead: 43s/patch \\

Pasini et al.~\cite{pasini2026evaluating} & 2026 & I & AD & RAG + Self-Ranking & XSS, SQLi & F2-score & LLM self-evaluation enhances detection robustness and reliability. \\

Singh et al.~\cite{singh2026multi} & 2026 & I & Smart Env. & SLM-as-a-judge & Evaluate quality of semi-synthetic WBSs. & Self-BLEU, Euclidean distance analysis (EDA) & DeepSeek-R1 achieved best balance of rubric-measured quality (> 92 \%), with good SELF-BLEU diversity score (< 6.5 \%).\\

Wang et al.~\cite{wang2026fine} & 2026 & I & CR-SE & HyVD-VP & Java enterprise code (Juliet 1.3) & Accuracy, F1-score & Hybrid pipeline; 95\% F1, identifies 16 previously unknown vulnerabilities. \\

\hline
\end{longtable}
}

\subsubsection{Attack Detection}

The use of LaaJ as an evaluation instrument for attack detection in the context of LLM represents a usage shift from traditional signature and statistics-driven approaches toward semantic, context-aware reasoning over heterogeneous security artifacts. In this setting, LLMs are not merely classifiers but act as judges that interpret, contextualize, and evaluate potential threats, often incorporating multiple sources of information such as code, logs, knowledge graphs, and external intelligence. Recent work demonstrates that LLMs can operate across diverse threat models, including software supply chain attacks, social engineering, and network intrusions, while also enabling explainability and adaptive reasoning capabilities that are difficult to achieve with conventional systems.

The authors of~\cite{zahan2025leveraging} focus on detecting malicious npm packages in the software supply chain. In this setting, the LLM is used as a \textit{primary semantic judge}, evaluating whether a package is malicious by reasoning over code, metadata, and dependency structures. They propose \textit{SocketAI}, which integrates traditional static analysis with LLM-based reasoning in a multi-stage pipeline, where heuristic filtering reduces the analysis space before invoking an LLM for deeper inspection. Unlike rule-based approaches that rely on predefined signatures, the LLM infers latent malicious intent, such as data exfiltration or unauthorized execution. This semantic abstraction allows the system to generalize to previously unseen attack patterns, effectively positioning the LLM-as-a-Judge that evaluates behavior based on inferred intent rather than explicit rules. However, despite achieving a low misclassification rate, challenges such as mode collapse and hallucination persist, along with limitations in handling large files and susceptibility to prompt injection.

In contrast to direct detection, the work~\cite{webb2025synthetic} pivots from direct classification to LLM-driven simulation of adversarial scenarios, leveraging generative capabilities to craft realistic phishing and social engineering narratives conditioned on user roles and organizational contexts. Here, the LLM plays an \textit{indirect judging role} by generating adversarial scenarios that define the evaluation space. Instead of performing explicit classification, the model produces realistic phishing and manipulation narratives conditioned on user roles and organizational contexts. These scenarios are then used to evaluate human detection capabilities, where LLM-as-a-Judge metrics assess generation quality while humans perform the final detection. Llama 3.1 with Chain-of-Thought prompting demonstrates strong capability in generating coherent and persuasive attacks, though challenges remain in ensuring realism across edge cases and avoiding over-generation of implausible scenarios.

The paper~\cite{belcastro2025enhancing} introduces a novel methodology, called \textit{KLAGE}, that combines Knowledge Graphs (KGs), eXplainable AI (XAI) techniques, and LLMs to provide a comprehensive solution for network security monitoring, threat analysis, explainability, and automated report generation. In this framework, the LLM functions as a \textit{context-aware reasoning judge}, operating over knowledge graph representations to evaluate potential threats. The LLM consumes graph-derived context and performs reasoning over relational structures, producing both threat classifications and natural language explanations. This hybrid approach effectively elevates the LLM to a judge over structured evidence, combining the precision of symbolic representations with the flexibility of neural reasoning. The resulting system not only improves detection performance but also enhances interpretability, as the LLM can articulate the rationale behind its decisions. The effectiveness of this approach is tightly coupled to the completeness and quality of the underlying knowledge graph, and scalability remains a concern in dynamic, large-scale environments.

A supporting direction is presented in the work~\cite{blefari2025cyberrag}, which extends traditional retrieval-augmented generation by introducing a modular, agent-based architecture for real-time attack detection and reporting called \textit{CyberRAG}. In this setting, the LLM acts as a \textit{knowledge-grounded judge}, refining its decisions through iterative retrieval and reasoning. Unlike standard RAG pipelines that often retrieve irrelevant context and lack justification capabilities, CyberRAG employs a central LLM agent that orchestrates multiple specialized components, including fine-tuned classifiers tailored to specific attack families, tool adapters for enrichment and alerting, and an iterative retrieval and reason loop over a domain-specific knowledge base. This agentic setup enables adaptive reasoning and flexible control flow, where new knowledge sources or components can be integrated without retraining the core LLM. Empirical evaluation demonstrates that this approach achieves high decision quality and robust, interpretable outputs, highlighting its potential for real-time, knowledge-grounded evaluation in complex environments

he paper~\cite{farrukh2025xg} proposes the \textit{XG-NID framework}, a hybrid intrusion detection system that combines graph neural networks (GNNs) with \textit{LLM-based reasoning}. Network traffic is modeled as a graph, where nodes represent entities (e.g., hosts, flows) and edges capture interactions, allowing the GNN to learn structural and temporal patterns of intrusion behavior. The LLM is employed as a \textit{post-hoc evaluator}, interpreting the GNN’s predictions, providing semantic context and generating human-readable explanations for detected anomalies. Beyond interpretation, the LLM also assesses the quality of its own explanations, linking clarity and correctness to improved classification reliability. This dual role enables the system to integrate low-level structural insights with high-level semantic reasoning, enhancing both detection performance and interpretability, though it introduces additional system complexity.


The study~\cite{pasini2026evaluating} reframes LLM-based attack detection as a generation-and-evaluation problem, focusing on improving robustness against injection attacks such as Cross-Site Scripting (XSS) and SQL injection (SQLi). In this framework, the LLM operates as a \textit{self-judge}, generating multiple candidate detection strategies via Retrieval-Augmented Generation (RAG) and then selecting the most robust option based on internal consistency through a Self-Ranking mechanism. RAG grounds the detector in external security knowledge, while Self-Ranking allows the LLM to assess the quality and reliability of its own outputs. This dual role—generation plus self-evaluation—enhances detection reliability, ensuring that selected detectors are both robust and consistent. Empirical results demonstrate substantial improvements in metrics such as F2-score, highlighting the value of integrating knowledge grounding with LLM-driven self-assessment, although the approach introduces additional computational cost and depends on the quality of the retrieved knowledge. Within the broader context of LLM-as-a-Judge, this work is notable because it implements self-judgment and robustness selection inside the detection pipeline itself, rather than relying on external evaluation. 

Finally, a work~\cite{shao2025effective} investigates agentic LLM systems for offensive security tasks, particularly in Capture-the-Flag (CTF) challenges, where the LLM functions as a \textit{trajectory-level judge}. The proposed \textit{CTFJudge} framework evaluates both intermediate reasoning steps and final outcomes of agent actions, providing fine-grained assessment of multi-step attack strategies rather than relying solely on binary success signals. To quantify partial correctness, the authors introduce the CTF Competency Index (CCI), which measures how closely an agent’s solution aligns with human-crafted ground truth. The study also examines the influence of LLM hyperparameters (e.g., temperature, top-p, token limits) on agent behavior, highlighting their role in balancing exploration and reliability. Supported by the CTFTiny benchmark for reproducible experimentation, this work extends the notion of LLM-as-a-Judge beyond final decisions to process-aware evaluation, assessing the quality of reasoning and intermediate actions in complex tasks.

Taken together, these developments illustrate a progression from standalone semantic detection to integrated, multi-component evaluation systems, where LLMs act as central, context-aware decision-making entities. Beyond generating outputs—such as detection logic, explanations or attack strategies—LLMs increasingly \textit{assume the role of evaluators}, assessing the quality, consistency, and reliability of their own outputs. By iteratively reasoning over retrieved knowledge, intermediate results, or multi-step processes, they can refine predictions, provide interpretable explanations, and maintain robustness against complex or previously unseen scenarios. This shift transforms \textit{LLMs into adaptive, process-aware evaluators}, enhancing trustworthiness and interpretability across diverse applications, though it also introduces challenges related to computational overhead, scalability, and dependence on high-quality external knowledge. Collectively, these trends underscore the need for principled approaches to designing, validating, and deploying LLM-based evaluation systems in security settings.

\subsubsection{Code Review and Software Engineering}
In a typical Software Development Life Cycle (SDLC), software systems progress through several stages, including planning, requirement definition, design, implementation, testing, deployment, and maintenance. Throughout this process, various monitoring and evaluation mechanisms are applied to ensure visibility into the development progress, maintain quality control, manage risks, and support reliable cost estimation. Among these mechanisms, evaluation tasks such as code review, patch validation, testing report assessment, and vulnerability analysis play a critical role in determining whether software artifacts meet the requirements necessary for integration into production systems. Recent research has increasingly explored the use of LLMs to assist or automate these evaluation activities within software engineering pipelines.

\begin{figure}[t]
    \centering
\tikzset{
    basic/.style  = {draw, text width=3cm, align=center, fill=gray!60, font=\sffamily, rectangle, rounded corners=2pt,text width=2.5cm},
    onode/.style = {basic, thin, rounded corners=2pt, align=left, fill=gray!10, text width=3cm},
    tnode/.style = {basic, align=left, fill=white, text width=4cm},
    xnode/.style = {basic, thin, rounded corners=2pt, align=center, fill=gray!30}
}

\begin{forest}
for tree={
    grow=east,
    growth parent anchor=west,
    parent anchor=east,
    child anchor=west,
    edge path={
        \noexpand\path[\forestoption{edge},->,>={latex}]
        (!u.parent anchor) -- +(5pt,0pt) |- (.child anchor)
        \forestoption{edge label};
    },
    font=\sffamily
}
[Code Review and\\Software Engineering, basic, l sep=10mm
    [Governance of LLM-Generated Code ~\cite{baek2026llm}, tnode]
    [Code Quality and Vulnerability Analysis \cite{wang2026fine,morales2025impromptu}, tnode]
    [Evaluation of Software Artifacts ~\cite{wang2025multi,zhou2025se,shi2025humanin,ji2026codegen}, tnode]
    [LLM-Assisted Code Review ~\cite{goldman2025types,olewicki2026impact,sghaier2025harnessing,jaoua2025combining}, tnode]
]
\end{forest}
    \caption{Taxonomy of LLM-based research directions in code review and software engineering.}
    \label{fig:code_review_taxonomy}
\end{figure}
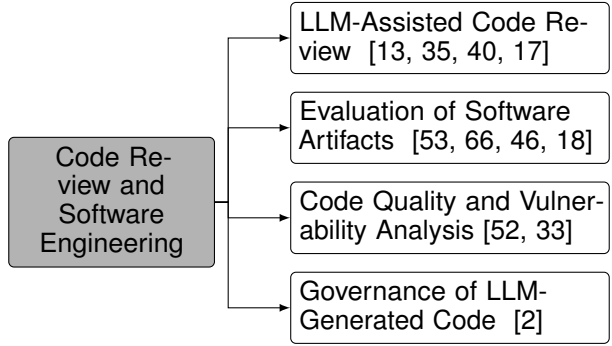

During our review process, we identified several contexts in which LLMs are used as automated judges within software engineering pipelines. Figure~\ref{fig:code_review_taxonomy} presents the taxonomy used to organize this body of work. The taxonomy distinguishes four main categories: {\em(i)} \textit{LLM-assisted code review}, where models generate or recommend feedback during review processes; {\em(ii)} \textit{evaluation of software artifacts}, where models assess the quality or correctness of artifacts such as patches, testing reports, or generated programs; {\em(iii)} \textit{code quality and vulnerability analysis}, where LLMs assist in detecting defects or security issues in source code; and {\em(iv)} \textit{governance of LLM-generated code}, which focuses on identifying and managing artifacts produced by generative models.\\

\noindent
\textbf{LLM-Assisted Code Review.} Code review remains one of the most important stages in the software development life cycle. However, it is extremely cognitively demanding and often becomes a bottleneck in the development process. During our review, we identified several works that explore how LLMs can support code reviewers by automatically generating or recommending feedback on submitted patches. For example, Olewicki et al.~\cite{olewicki2026impact} propose \textit{RevMate}, an LLM-based review assistant. The core motivation of their work comes from the observation that prior LLM-based automation studies rely on offline benchmarks using synthetic datasets. Authors of this paper investigate whether LLM-generated review comments can provide practical value to developers in realistic industrial environments. \textit{RevMate} is deployed within the habitual review workflows of professional engineers and evaluates whether its suggestions are acted upon and considered useful. \textit{RevMate} is built on GPT-4o and combines two complementary techniques: {\em(i)} Retrieval-Augmented Generation (RAG), which retrieves relevant code context and historical review examples to ground comment generation, and {\em(ii)} a LaaJ mechanism that filters low-quality outputs before presenting them to reviewers. They deployed \textit{RevMate} for six weeks across two organizations involving 59 reviewers, 587 patch reviews, and approximately $1,600$ generated comments evaluated in context. Their results show that acceptance rates were between $8.1\%$ and $7.2\%$ for the direct incorporation of generated comments. An additional $14.6\% -20.5\%$ of comments were considered valuable suggestions even when they were not accepted. Additionally, refactoring-related comments significantly outperformed functional comments (approximately $18\%$ versus $5\%$ acceptance). 

Building on this question of actionability, Goldman et al.~\cite{goldman2025types} investigate not just whether LLM-generated comments are acted upon, but which types of comments most reliably trigger code changes. Their motivation stems from the observation that not all generated review comments will drive code modifications, and that understanding this distinction is crucial for improving the practical effectiveness of automated review tools. To do so, they develop a five-category taxonomy, Readability, Bug, Maintainability, Design, and No Issue, and construct an LLM-as-a-Judge pipeline using GPT-4.1 to classify comments at scale, validated against human annotators. Their empirical study spans approximately $4,000$ LLM-generated comments from Atlassian internal projects and $1,000$ human-written comments from open-source repositories. Notably, this approach enables direct comparison between the two reviewer types. Their results show that readability, bug, and maintainability comments had significantly higher resolution rates than design-oriented ones. They also find that LLM reviewers and human reviewers exhibit complementary strengths: the LLM generates substantially more bug and maintainability comments, while humans tend toward broader design feedback. Limitations include the use of line-level commit proximity as a proxy for resolution, which may not always reflect genuine comment-driven changes, and a dataset skewed toward specific language ecosystems. 

The findings from~\cite{goldman2025types} raise a question: if certain comment types are empirically more actionable than others, are the datasets used to train these models actually reflecting that signal, or are they polluted with the very comment types developers tend to ignore? Sghaier et al.~\cite{sghaier2025harnessing} argue that the existing code review datasets are extracted raw from repositories and contain uncivil, verbose, unclear, and irrelevant comments, which causes models to learn undesirable patterns. They argue that curating the training data yields significant downstream gains compared to designing better model architectures. Their methodology starts by constructing an evaluation framework with three categorical dimensions (Type, Nature, Civility) and three scoring criteria (Relevance, Clarity, Conciseness). This framework is then used to audit the original dataset using Llama-3.1-70B as-a-Judge, which they also validate against human annotators using Cohen's kappa. The curation pipeline consists of two stages: first, filtering, where low-relevance comments are removed based on their scores, and second, reformulation, where the same LLM is used to improve the clarity of the remaining comments without changing their semantic intent. Their results show that fine-tuning on the curated dataset yields a $46\%$ improvement in BLEU for comment generation and a $22\%$ improvement in CodeBLEU for code refinement compared to training on the original dataset. These results show that in the context of code review automation, data quality can be a more impactful factor than model design choices.

Beyond data quality, another line of work questions whether LLMs alone are sufficient for code review regardless of how well they are trained. Jaoua et al.~\cite{jaoua2025combining} argue that in general automated code review faces a fundamental trade-off: rule-based static analyzers are precise but limited in coverage, while fine-tuned LLMs cover more ground but sacrifice precision. They propose combining both by injecting static analysis knowledge into an LLM pipeline at three different points. The first strategy augments training data with reviews from both sources, filtered by a judge model for quality. The second injects static analyzer outputs directly into inference-time prompts. The third simply concatenates the outputs of both systems after inference. Their results show that the prompt injection approach delivers the largest accuracy gain, the data augmentation approach achieves the strongest coverage, and the simple concatenation strategy yields only marginal improvements, and in fact inherits LLM errors and/or produces conflicting signals. Limitations include the study being restricted to Java code, the unresolved issue of LLM outputs sometimes contradicting static analyzer findings, and a potential bias from using the same LLM family for both generation and evaluation.

Taken together, all of these works trace a common line in LLM-assisted code review: generating a comment is easy, generating one a developer will trust, act on, and benefit from requires getting the comment type, the training signal, and the underlying tooling right simultaneously. RevMate~\cite{olewicki2026impact} and Goldman et al.~\cite{goldman2025types} establish that actionability varies significantly by comment type and context. Sghaier et al.~\cite{sghaier2025harnessing} show that the training data underpinning these models is itself a source of noise that curations can meaningfully address. Jaoua et al.~\cite{jaoua2025combining} further demonstrate that LLMs alone may be structurally insufficient for high precision review, and that hybrid integration with static analysis is a promising but unresolved direction.
\\

\noindent
\textbf{Evaluation of Software Artifacts.} In previous paragraphs, we examined works that assess the quality of LLM-generated code review comments. If LLMs exhibit systematic limitations in reviewing code, meaning struggling with actionability, precision, and grounding, a natural question arises: can LLMs reliably evaluate the quality of other software testing artifacts, such as the test reports that crowdsourced testers submit? This question has motivated a parallel line of work applying the LaaJ to structured software engineering deliverables whose quality is inherently multi-dimensional and whose manual assessment is both expensive and demonstrably subjective. 

Wang et al.~\cite{wang2025multi} tackle the problem of assessing crowdsourced testing reports at scale, where manual review by human experts is slow, costly, and (as they demonstrate) only moderately consistent at the individual rater level (ICC = $0.49-0.63$). They propose a multi-agent LLM framework that evaluates each report along three dimensions: textuality, which scores the structural and linguistic quality of individual test cases and defect reports; adequacy, which measures how well a report's test cases cover the functional requirements specified in the project documentation; and competitiveness, which quantifies how rare and unique a tester's discovered defects are relative to all other submissions in the same task. Each dimension is handled by a dedicated agent. The textuality agent decomposes scoring into binary checklists to avoid LLM ordinal instability and resolves disagreements between two independent models via a third arbitration step. The adequacy agent builds a hierarchical requirement tree from specification documents and maps test cases to atomic leaf nodes. The competitiveness agent clusters all defect reports using LLM-based hierarchical clustering and assigns inverse-frequency scores and rewards testers who uncover rarer defects.  They evaluate their proposal on 64 crowdworkers across three mobile applications, and according to their results, the framework achieves strong agreement with aggregated human judgments (QWK of $0.72-0.82$ for textuality, $0.89-0.97$ for adequacy) and delivers $5-90\times$ speedups over manual review. Limitations include the absence of direct human validation for the competitiveness dimension, the textuality checklist requiring domain-expert refinement per platform, and all experiments being confined to mobile application testing.

While Wang et al.~\cite{wang2025multi} demonstrate that LLM-based multi-agent evaluation can match human judgment on crowdsourced testing reports, their framework is purpose-built for that artifact type, with dimension definitions, checklist structures, and clustering logic all tailored to mobile application testing. A broader and arguably more fundamental question remains: can a unified LLM-as-a-Judge framework reliably assess the correctness of the diverse range of software artifacts that automated SE tools produce, spanning generated code snippets, bug-fixing patches, and code summaries, without requiring bespoke engineering for each artifact type? Zhou et al.~\cite{zhou2025se} tackle the problem of automatically evaluating the correctness of generated software artifacts, where similarity-based metrics such as BLEU~\cite{papineni-etal-2002-bleu}, which measures n-gram overlap between a generated output and a reference, and CodeBLEU~\cite{ren2020codebleumethodautomaticevaluation}, which extends BLEU with syntactic and data-flow matching to better account for code structure, are known to misalign with human judgment as they assess surface-level similarity rather than functional correctness. The representative LLM-as-a-Judge baseline, ICE-Score~\cite{zhuo-2024-ice}, improves over similarity-based metrics by prompting an LLM to directly assign a correctness score based on predefined criteria, but relies on a single fixed evaluation strategy that limits the diversity of reasoning perspectives. To address this, they propose SE-Jury, an LLM-as-Ensemble-Judge framework that defines five evaluation strategies: direct scoring, score-then-rethink with self-reflection, semantic equivalence comparison against a reference, reference decomposition into key correctness properties followed by per-property verification, and LLM-generated test case simulation. A dynamic team selection mechanism evaluates all valid strategy subsets on 20 annotated samples and selects the highest-correlation subset before scoring the full dataset. They evaluate SE-Jury across four human-annotated benchmarks spanning code generation, automated program repair, and code summarization over five programming languages, and according to their results, SE-Jury outperforms all baselines by $29.6\%-140.8\%$ in average correlation with human judgments, with human-tool agreement on automated program repair approaching inter-human agreement levels. Limitations include a persistent gap on code summarization and reliance on commercial LLMs.

Shi et al.~\cite{shi2025humanin} extend this direction by introducing a human-in-the-loop patch evaluation framework in which automated judgments are anchored to developer-refined rubrics (scoring tool) to improve reliability. Their motivation stems from an empirical observation that manual patch assessment, the de facto gold standard in Automated Program Repair (APR) evaluation, itself suffers from low inter-rater reliability (Fleiss's $\kappa = 0.31$) when raters work independently, yet agreement improves substantially when raters share a structured rubric (Cohen's $\kappa$ rising to $0.53-0.84$ across rater pairs). They propose a two-stage pipeline. In the first stage, an LLM generates a per-bug draft rubric from the bug description and ground truth patch, which is structured around a fixed template capturing root cause, fix requirements, and concrete examples of acceptable and unacceptable solutions; two human experts then review and refine this draft once per bug into a golden rubric. In the second stage, a separate LLM judge evaluates candidate patches against the golden rubric, outputting a binary valid/invalid label with a natural-language justification. Evaluated on 115 fail-to-pass patches across 48 sanitizer bugs from Google's internal monorepo, the framework achieves Cohen's $\kappa = 0.75$, precision $0.80$, and recall $0.94$ on the $70.4\%$ of patches where human raters unanimously agreed, with a negative predictive value of $0.94$, making the invalid predictions highly reliable. Ablation studies confirm that human rubric refinement is the dominant contributor: omitting it drops $\kappa$ from $0.57$ to $0.38$, and replacing the structured template with a free-form rubric degrades performance further to $\kappa = 0.29$. Limitations include restriction to a single proprietary codebase and bug class, reliance on Gemini 2.5 Pro throughout, and a remaining precision gap ($\kappa = 0.57$) on contested patches where human raters themselves disagree.

Ji et al.~\cite{ji2026codegen} extend the judging paradigm to a qualitatively different artifact type: executable code whose output is assessed not by functional correctness but by perceptual and semantic fidelity to a natural language specification. They introduce CodeGen-3D, a benchmark for evaluating LLMs that generate Blender Python (bpy) scripts for 3D object modeling, a domain where no prior standardized evaluation existed. The framework evaluates generated artifacts along three axes: execution reliability, measured as the fraction of scripts that fail to compile or run in headless Blender with a typed taxonomy distinguishing syntax, API, and runtime failures; semantic alignment, computed as average and maximum CLIP similarity between six orthographic renderings and the input prompt; and human-aligned perceptual quality, assessed via GPT-4o in a pairwise preference task comparing the best generated view against the best ground-truth view. Applied to ten systems, eight general-purpose LLMs, a fine-tuned BlenderLLM, and an iterative GPT-4.1 generator paired with a GPT-4o critic, the benchmark reveals a sharp execution reliability gap: specialized and iterative systems achieve $3-4\%$ failure rates while general LLMs range from $21\%$ to $92\%$. A key methodological finding is that CLIP scores cluster tightly across models even when GPT-4o preferences diverge substantially, showing that static text image similarity significantly underestimates perceptual quality differences and motivating multi-signal evaluation. Limitations include a small prompt set (100 single-object prompts), reliance on a single VLM judge, and the absence of structure-aware geometric metrics. For this body of work, CodeGen-3D is notable as an instance of the judging paradigm being applied to generative code artifacts outside classical SE domains. It illustrates both the generality of the LLM-as-a-Judge approach and the persistent need to complement automated metrics with human-aligned preference signals.
\\

\noindent
\textbf{Code Quality and Vulnerability Analysis.} The previous section has shown that LLMs can reason about code semantics well enough to review testing artifacts; what about whether they can reliably detect real security vulnerabilities? Wang et al.~\cite{wang2026fine} address this by proposing HyVD-VP, a hybrid vulnerability detection pipeline targeting Java enterprise codebases where traditional static analyzers such as Fortify and SpotBugs achieve F1-scores below $79\%$. Their approach fine-tunes a Qwen3-8B model via supervised fine-tuning (SFT) followed by direct preference optimization (DPO), the first application of DPO in this domain, then segments long source files into semantically coherent slices via control flow graph traversal, enriching each slice with retrieved vulnerability knowledge at inference time. A PAL-based module further generates sandboxed runtime test scripts for flagged slices, and an LLM-as-a-Judge layer reconciles the static and dynamic signals into a final verdict. Evaluated on a custom long-code Java dataset and the Juliet Java 1.3 benchmark, HyVD-VP achieves $95.2\%$ accuracy and $95.1\%$ F1, outperforming Fortify by $18.3\%$ in F1 and surpassing prompt-only LLM baselines by over 20 percentage points, while identifying 16 previously unknown vulnerabilities across six enterprise projects.

Morales et al.~\cite{morales2025impromptu} address a complementary challenge: the lack of principled abstractions for defining, managing, and migrating prompts across generative AI platforms. They introduce Impromptu, a model-driven engineering (MDE) framework comprising a platform-independent domain-specific language (DSL) and accompanying toolchain. Prompts are formalized as typed assets with structured snippet components (prefix, core, suffix), multimodal inputs, hyperparameter bindings, versioning metadata, and support for prompt chaining, all governed by OCL well-formedness constraints. Code generators then transform these abstract specifications into platform-specific prompts and executable Python stubs targeting OpenAI (ChatGPT, DALL-E 3) and Stable Diffusion/Midjourney, handling platform divergences such as attention weighting syntax and negative prompt semantics. An LLM-as-a-Judge validation layer automatically generates trait-level test queries to assess response compliance at runtime. Applied to the LangBiTe bias-testing library, Impromptu collapsed 207 duplicated formatting snippets into single reusable definitions, substantially reducing maintenance overhead. Limitations include a single qualitative case study with no quantitative cross-platform migration benchmarks, narrow target system coverage, and inherited reliability concerns from the LLM-as-a-Judge evaluation paradigm.\\

\noindent \textbf{Governance of LLM-Generated Code.} The preceding pargraphs examined LaaJ in settings where the code under evaluation is assumed to be human-authored and the judge's role is to assess its quality, correctness, or security. A complementary challenge arises when the provenance of the code itself is in question, that is, when the LaaJ must determine not what the code does, but whether it was produced by an LLM in the first place. This shift from evaluating code artifacts to governing their origin introduces distinct reliability concerns, particularly in educational contexts where student skill development is at stake. Baek et al.~\cite{baek2026llm} address this by proposing an explainable detection framework for identifying LLM-generated code in introductory Python programming courses, where prior approaches such as Hoq et al.~\cite{10.1145/3626252.3630826} achieved over $97\%$ accuracy through binary classification but offered no interpretability (no interpretability is a critical limitation in settings where transparent, actionable feedback is pedagogically essential). Their pipeline proceeds in three stages: GPT-4o and seven faculty instructors collaboratively identify four discriminative features per code class (e.g., inconsistent variable naming and manual indexing for student-written code; descriptive identifiers and defensive edge-case handling for LLM-generated code), which are then used as prompting scaffolds to automatically annotate a $4{,}000$-sample dataset with natural language explanations. Three code-specialized LLMs of approximately 7B parameters ( CodeLlama, CodeGemma, and DeepSeek Coder) are subsequently fine-tuned via LoRA to jointly predict a binary label and generate a supporting explanation in a single forward pass. Their ablation study shows that including explanation supervision is critical: omitting it causes the LLM-as-a-Judge explanation score to collapse from $4.90$ to $2.43$ out of $5$, while detection accuracy degrades only marginally by $0.5\%$. These results prove that explanation quality is not a free byproduct of classification training. The best-performing model, CodeGemma-FT, achieves $99.3\%$ detection accuracy, which it surpasses human instructors by $12.8$ percentage points ($86.5\%$), and attains an explanation score of $4.90$ under automated evaluation and $4.24$ under human assessment. When applied retrospectively to course submissions spanning 2022--2024, their detector reveals a sharp rise in LLM-generated submissions, from nearly $0\%$ in Autumn 2022, prior to the release of ChatGPT, to over $40\%$ in Autumn 2024 for the most challenging exercises, this validates both the detector's temporal consistency and the real-world urgency of the governance problem. Their limitations include confinement to five exercises from a single institution, a naive prompting strategy that may not reflect diverse real-world student LLM usage patterns, and a persistent false positive risk for high-performing students whose concise, idiomatic code stylistically resembles LLM output. \\

\noindent Across all these papers we looked into, we can easily observe that the LLM-as-a-Judge paradigm has become a crucial component in software engineering evaluation pipelines. It is serving roles which range from filtering low-quality review comments and scoring crowdsourced test reports to reconciling static and dynamic vulnerability signals and validating prompt compliance at runtime. However, the very properties that make LaaJ attractive in these settings, namely: its scalability, its capacity for nuanced semantic reasoning, and its ability to replace slow and inconsistent human assessment also amplify the security concerns. Several of the  works we surveyed explicitly acknowledge that using the same LLM family for both generation and judgment introduces circular evaluation risks~\cite{jaoua2025combining}, that LLM judges can be misled by surface-level plausibility in code artifacts~\cite{ji2026codegen, zhou2025se}, and that the reliability of the entire pipeline hinges on structured human oversight that is difficult to scale~\cite{shi2025humanin}. Moreover, when LaaJ is embedded in security-critical workflows, the consequences of an unreliable judge extend well beyond degraded evaluation quality. To illustrate, in vulnerability detection~\cite{wang2026fine}, a faulty judge can introduce false assurances that mask real security flaws, while in academic integrity enforcement~\cite{baek2026llm}, it can misclassify high-performing students whose idiomatic code stylistically resembles LLM output, a false positive problem that the authors themselves caution against addressing through direct penalization. These are only two instances among the works surveyed in this section, but they underscore a broader theme of our systematization: as LaaJ transitions from an offline evaluation convenience to an inline decision-making authority in production software systems, the attack surface it exposes and the trust it demands both grow commensurately, making the robustness and adversarial resilience of LLM-based judges an increasingly urgent research priority.

\subsubsection{Smart Environment}


Several works have focused on the applicability of LaaJ to diverse areas of the Smart Environment domain, including water treatment~\cite{xu2025towards}, banking~\cite{stavarache2025enhancing}, construction schedules~\cite{singh2026multi}, and other industrial settings~\cite{krayko2025rurage}.

\begin{figure*}[ht]
    \centering
\begin{tikzpicture}[scale=2, every node/.style={scale=0.9}]

\draw[->, thick] (0,0) -- (6,0) node[right]{Flexibility};

\node[below] at (0,0) {Low};
\node[below] at (6,0) {High};

\filldraw (0.5,0) circle (2pt);
\node[above, align=center, rotate=0] at (0.5,0.1) {Keyword-based\\(e.g., BLEU, ROUGE)};

\filldraw (2,0) circle (2pt);
\node[above, align=center, rotate=0] at (2,0.1) {Semantic-based\\(e.g., BERT)};

\filldraw (3.5,0) circle (2pt);
\node[above, align=center, rotate=0] at (3.5,0.1) {Hybrid\\LaaJ + Others};

\filldraw (4.75,0) circle (2pt);
\node[above, rotate=0] at (4.75,0.1) {LaaJ};

\filldraw (5.75,0) circle (2pt);
\node[above, align=center, rotate=0] at (5.75,0.1) {Human\\Annotator};

\end{tikzpicture}
\caption{LaaJ positions itself as a middle-ground between the expensive but flexible human annotation and cheaper but more rigid methods of evaluation.}
\label{fig:laaJOnContinuum}
\end{figure*}
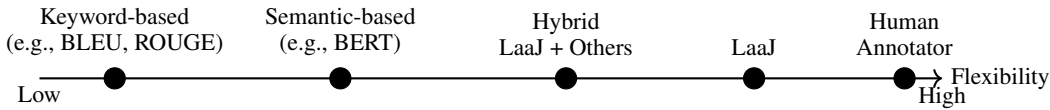

A common theme is the emergence of LaaJ as a middle-ground between the expensive (but highly flexible) human-annotator approach and the cheaper (but more rigid) techniques to evaluate LLM outputs, such as keyword-based methods (e.g., BLEU, ROUGE) and semantic-similarity based methods (e.g., BERT). LaaJ and other methods are depicted as points on a flexibility axis continuum in figure \ref{fig:laaJOnContinuum}.

If used well, LaaJ presents the opportunity to reduce the workload of human domain experts who have to perform data annotation, allowing them to focus on other tasks. Importantly, LaaJ should be designed in a way that these human annotators can easily review its work and steer it in the right direction (e.g., by contributing to designing the judge's prompts and rubrics, or by providing examples of valid judgments), as the LLMs used for LaaJ may lack the necessary domain-specific knowledge to produce high-quality judgments if left unchecked. LaaJ should be seen as a way to augment the potential of a limited pool of individual experts, rather than as an infallible replacement for their judgment.

In a number of Smart Environment scenarios, using a frontier/cloud model to evaluate LLM output can raise concerns about privacy and/or operating expenses, which can be a critical obstacle to the adoption of LaaJ.  Some LaaJ schemes may require an excessive number of remote API calls, whose cost may not be entirely justified by the use case.  Sometimes, evaluating a model's response may require expert knowledge that a small organization may not be willing to share with cloud providers.

For cases like these, the use of Small Language Models (SLMs) as Judges (or SaaJ) has been explored. SLMs are (informally) defined as language models with a parameter count in the ballpark of a few tens of billions of parameters (e.g., 1B, 7B, 14B, 20B). These models can be quantized to run efficiently on consumer hardware, in contrast to bigger models that require server-grade infrastructure and should be accessed via cloud services. 

SaaJ has the potential to increase privacy and reduce inference costs. The tradeoff, compared to cloud LLMs, is a decreased amount of world knowledge embedded in parametric memory and a weaker reasoning ability. To mitigate these limitations of SLMs, techniques such as ensembling, RAG, and finetuning can be explored. Ultimately, the works that we explored emphasize the necessity to develop optimized LaaJ techniques that may differ from domain to domain, avoiding over-reliance on generic toolkits whose LaaJ-powered metrics may present skewed results when applied to arbitrary fields.

\subsubsection{Other Application Domain}


LaaJ has been applied to a wide variety of other domains, such as evaluating the quality of generic Q\&A systems~\cite{balog2025rankers}, \cite{abeyratne2025alignllm}, benchmarking the performance of RAG pipelines in the legal~\cite{li2025lexrag} domain, and in evaluating LLM-based recommender systems~\cite{deldjoo2025toward}.

LaaJ is increasingly applied to security-related scenarios, such as: computing Attack Success Rate (ASR) from prompt injection attacks~\cite{wang2025manipulating}, assessing a model's demographic bias or toxicity~\cite{cantini2025benchmarking}, estimating the confidence of a model's output before it is generated~\cite{shen2025defining}, locally redacting sensitive user data before it is sent to more powerful LLMs on the cloud~\cite{zhou2025rescriber}, or judging the effectiveness of hybrid local-cloud LLM data processing pipelines~\cite{zhang2025lsrp}.

Several mechanisms fall under the umbrella of LaaJ: from simple pointwise evaluation to pairwise and listwise approaches. Some approaches make use of output token logits to soften a model's prediction, thus requiring ``white-box" access to the judge model. Ensembling techniques have also been explored, such as averaging the weighted outputs of several rubrics (i.e. structured scoring guides) that cover different aspects to be evaluated, aggregating judgments from several LLMs (usually SLMs) with a voting scheme~\cite{singh2026multi}, using embedding-based similarity techniques with hypothetical questions and answers~\cite{abeyratne2025alignllm}, or averaging several weaker metrics into a combined overall score.
Some approaches integrate external (non-parametric) knowledge into LaaJ using RAG mechanisms~\cite{abeyratne2025alignllm}.

A subtle issue that emerged in some of these works relates to the use of meta-prompting to assess the adequacy of a rubric and to reveal potential problems in the evaluation scheme~\cite{cantini2025benchmarking}. This suggests the need to develop formal frameworks that can make effective use of ``higher-order judges" without spiraling into infinite regress.

The integration of LaaJ into safety-critical systems underscores the importance of knowing and mitigating the various forms of bias that may arise from its usage. Common forms of bias include (among others): re-ranker bias, preference bias, sentiment bias, token bias, verbosity bias, and position bias.

Re-ranker bias manifests as a preference for a judge model for LLM-based re-rankers, possibly explained by the structural similarity between many LaaJ tasks and LLM-based candidate re-ranking techniques~\cite{balog2025rankers}.

Preference bias (or self-enhancement bias) ~\cite{abeyratne2025alignllm} is the preference of an LLM for its own outputs; in multi-model LaaJ ensembling scenarios, this can be mitigated by the ``leave-one-out strategy": when judging the output of model $M$, leave it out from the pool of judges and use all other models instead~\cite{cantini2025benchmarking}. Verbosity bias is the preference of LaaJ for longer outputs that are not necessarily more complete or correct. This can be mitigated by introducing a length penalty to the LaaJ's score~\cite{abeyratne2025alignllm}. Position bias, possibly related to the lost in the middle phenomenon~\cite{liu2023lostmiddlelanguagemodels}, applies to pair-wise and list-wise LaaJ evaluation modes. Position bias causes the LLM to judge a candidate based on its position in the prompt rather than its intrinsic quality. This can be mitigated (at the expense of longer processing times) by performing multiple rounds of evaluation, swapping the positions of the candidates at each round~\cite{abeyratne2025alignllm}. Token bias is a related failure mode, where an LLM may treat certain tokens as ``semantic shortcuts'' and skip a reasoning step that may lead it to a different conclusion.

In addition to these biases, LaaJ may exhibit other important limitations, such as their inability to discern between subtle but meaningful performance differences across the evaluated systems~\cite{balog2025rankers}, their vulnerability to keyword stuffing (related to token bias) and to the wider adversarial prompting techniques such as prompt injection and jailbreaking, and their lack of domain-specific knowledge that is often needed to act as a judge in specific domains.

Ultimately, choosing the right LaaJ technique comes down to: {\em(i)} the specific evaluation task at hand, as well as to {\em(ii)} the careful formulation of a threat model and the analysis of possible biases and failure modes. Evaluation tasks that require judging different methods relative to each other may benefit from pairwise evaluation, while tasks that require the evaluation of each method against an absolute scale may profit from a pointwise scheme; and one must keep in mind the biases accrued by each of these methods. Similarly, cases where adversarial prompting is expected from the evaluated systems may require complementing LaaJ with some discriminative probing mechanism to ensure that the judge is not being cheated. 

To decide on a suitable LaaJ strategy for a given application in an arbitrary engineering domain, a starting point may be to consider the set of simple binary distinctions shown in \ref{tab:laajDimensions}. This table summarizes seven different independent dimensions of LaaJ; for each dimension, it compares the ``dominant paradigm'' to the ``alternative paradigm'', the latter being the less popular approach in the set of papers surveyed in the last two sections. The seven dimensions that were identified are the following:

\begin{itemize}
    \item \textbf{Model access}. This dimension refers to whether black-box access to the model used for LaaJ is enough or whether white-box access is required by the methods.
    \item \textbf{Scoring method}. This dimension can be absolute (i.e., each candidate is evaluated on its own, such as in pointwise methods) or relative (i.e., candidates are compared to each other, such as in pairwise or listwise methods).
    \item \textbf{Deployment}. Deployment of the LaaJ model can be done on the cloud or locally; as aforementioned, cost and privacy-related concerns play a role in choosing a deployment strategy.
    \item \textbf{Subject of evaluation}. This dimension refers to whether the LaaJ is being used to evaluate the performance of a model on some task (i.e., first-order LaaJ) or whether it is being used to evaluate another judge (i.e., higher-order LaaJ). Most works use first-order LaaJs only; some works have used higher-order LaaJ via meta-prompting (i.e., using an LLM to evaluate a prompt to another LLM).
    \item \textbf{Rubric design}. This dimension refers to whether multiple rubrics are ensembled or whether a single rubric is used on its own.
    \item \textbf{Knowledge source}. This dimension refers to whether the LaaJ model only has access to parametric memory (i.e., internal knowledge source), or whether it is given access to external sources of knowledge, for example, via sets of golden answers in a LaaJ-powered benchmark.
    \item \textbf{Usage stage}. This dimension refers to when LaaJ is applied with respect to the generation process. In the dominant paradigm, LaaJ is used in a \emph{post hoc} manner, i.e., to evaluate outputs after they have been generated. In the alternative paradigm, LaaJ is used \emph{a priori}, i.e., to guide or influence the generation process itself.
\end{itemize}

\begin{table*}[t]
\centering
\small
\begin{tabular}{p{2cm}|p{5.5cm}p{5.5cm}}
\hline
\textbf{Dimension} & \textbf{Dominant Paradigm } & \textbf{Alternative Paradigm } \\
\hline

Model access 
& Black box (text as output) \cite{stavarache2025enhancing} \cite{deldjoo2025toward} \cite{balog2025rankers} \cite{xu2025towards} \cite{zhang2025lsrp} \cite{singh2026multi} \cite{zhou2025rescriber} \cite{li2025lexrag} \cite{cantini2025benchmarking} \cite{krayko2025rurage} \cite{shen2025defining}

& White box (logits as output) Mentioned in~\cite{balog2025rankers} \\

Scoring method 
& Absolute/pointwise (assigns independent scores)  
\cite{stavarache2025enhancing} \cite{wang2025manipulating} \cite{deldjoo2025toward}, \cite{balog2025rankers}, \cite{xu2025towards} \cite{zhang2025lsrp} \cite{singh2026multi} \cite{zhang2025lsrp} \cite{singh2026multi} \cite{li2025lexrag} \cite{cantini2025benchmarking}  \cite{abeyratne2025alignllm} \cite{krayko2025rurage} \cite{shen2025defining}

& Relative / pairwise or listwise (ranks items) \cite{deldjoo2025toward} \cite{balog2025rankers}  \cite{zhou2025rescriber} \\

Deployment 
& Cloud (runs remotely) \cite{wang2025manipulating} \cite{deldjoo2025toward} \cite{balog2025rankers} \cite{zhou2025rescriber} \cite{shen2025defining}

& Local (runs on premise) \cite{balog2025rankers} \cite{xu2025towards} \cite{zhang2025lsrp}, \cite{singh2026multi}, \cite{abeyratne2025alignllm} \\

Subject of Evaluation
& First-order (judge a model) \cite{stavarache2025enhancing} \cite{wang2025manipulating} \cite{deldjoo2025toward} \cite{balog2025rankers} \cite{xu2025towards} \cite{zhang2025lsrp} \cite{singh2026multi} \cite{zhou2025rescriber} \cite{li2025lexrag} \cite{cantini2025benchmarking} \cite{abeyratne2025alignllm} \cite{krayko2025rurage} \cite{shen2025defining} 

& Higher-order (judge a judge) Used in~\cite{cantini2025benchmarking} (as meta-prompt) \\

Rubric design 
& Ensembling (aggregate multiple criteria) \cite{stavarache2025enhancing} \cite{xu2025towards} \cite{zhang2025lsrp} \cite{singh2026multi} \cite{li2025lexrag} \cite{cantini2025benchmarking} \cite{abeyratne2025alignllm} \cite{krayko2025rurage} 
& Single rubric (one criterion) \cite{wang2025manipulating} (but averages several runs), \cite{zhou2025rescriber}\\

Knowledge source 
& Internal (using model weights, or for logical consistency) \cite{stavarache2025enhancing}, \cite{wang2025manipulating} (attack, no attack), \cite{deldjoo2025toward}, \cite{balog2025rankers}, \cite{xu2025towards}, \cite{zhang2025lsrp}, \cite{singh2026multi}, \cite{zhou2025rescriber}, \cite{cantini2025benchmarking} (but uses builtin LLM world knowledge), \cite{krayko2025rurage}, \cite{shen2025defining} 

& External (injecting domain-specific knowledge) \cite{deldjoo2025toward},  \cite{xu2025towards}, \cite{zhang2025lsrp}, \cite{singh2026multi}, \cite{li2025lexrag} (LaaJ has access to golden answer), \cite{abeyratne2025alignllm} \\

Usage stage 
& Post hoc (evaluate generated output) \cite{stavarache2025enhancing} \cite{wang2025manipulating} \cite{deldjoo2025toward} \cite{balog2025rankers} \cite{xu2025towards} \cite{singh2026multi} \cite{zhou2025rescriber} \cite{li2025lexrag} \cite{cantini2025benchmarking} \cite{abeyratne2025alignllm} \cite{krayko2025rurage}
& A priori (drive generation) \cite{zhang2025lsrp}, \cite{shen2025defining} \\
\hline

\end{tabular}
\caption{Some high-level binary dimensions that can be used to broadly classify LaaJ systems. Each dimension is orthogonal to the others.}
\label{tab:laajDimensions}
\end{table*}

\section{Challenges and Open Problems}
\label{sec:challenge}

Although LLM-as-a-Judge systems offer significant opportunities for automating evaluation tasks, in practice, they face numerous security and reliability issues. Many of these issues arise from the inherent characteristics of Large Language Models. For example, LLM-based evaluation systems are highly sensitive to prompt manipulation, bias, and variability in outputs. So, generating reliable and secure evaluation frameworks requires addressing these issues. In this session, we discuss open research problems and security issues related to LLM-based judging systems.

\subsection{Vulnerability to Adversarial Prompt Manipulation }
 One of the most critical concerns in LaaJ is their susceptibility to adversarial prompt manipulation. Malicious users may evaluate a model's reasoning patterns and craft responses specifically designed to influence the decision-making process. Such responses include persuasive language, misleading explanations, or formatting methods that can bias the decisions of judges.LLMs depend heavily on contextual cues within prompts, so adversarial inputs manipulate evaluation outcomes. As a result, malicious outputs may receive high evaluation scores. Additionally, prompt injection attacks include hidden instructions within the candidate response, which can deviate the judging model from the correct evaluation criteria. These vulnerabilities raise serious concerns about the integrity and security of automated evaluation pipelines.

The open research problems in this context are: 

\begin{itemize}
    \item Create evaluation systems based on LLM  to prevent adversarial attacks.
    \item Design methods for detecting prompt injection and evaluation manipulation attacks.
    \item Design defence mechanisms for secured evaluation pipelines.
\end{itemize}

\subsection{Positional Bias and Evaluation Manipulation}
Positional bias is a common issue in LLM-based judging systems. The order in which candidate responses appear can influence the evaluation result. For example, in pairwise comparison tasks, the first response in the prompt may be evaluated more favorably by the judge model without considering its actual quality. This behaviour occurs because LLMs rely on contextual patterns. Such bias can create chances for evaluation manipulation. Attackers may intentionally exploit the ordering effect by strategically placing their responses in positions that increase the likelihood of being selected as a better answer. Even though to resolve this, suggested methods like response-order swapping, in automated evaluation frameworks, positional bias is one of the challenges. As a result, the fairness and reliability of automated evaluation systems may be compromised.

The open research problems in this context are: 

\begin{itemize}
    \item Design a positional-invariant evaluation framework.
    \item Develop a robust ranking mechanism for evaluating candidate responses.
    \item Quantify and reduce positional bias in evaluation tasks.
\end{itemize}

\subsection{ Length and Style Bias Exploitation}
Another significant challenge in LLM-based judging systems is length bias. Here, LLM judges tend to favor responses that are longer in length. Such responses often appear more descriptive and informative; however, the actual quality may be low. Despite this,  verbose answers receive higher evaluation scores simply due to their length. Like length bias, various stylistic features such as fluency, formatting, and rhetorical structure can influence evaluation responses independently of factual correctness. As a result, these responses that are well written may be highly rated, even if they contain inaccuracies. Adversaries can exploit these weaknesses by increasing the length of the response or stylistic presentation to manipulate the results. This undermines the fairness and reliability of LLM-based evaluation systems.

The open research problems in this context are: 

\begin{itemize}
    \item Develop evaluation metrics that give priority for factual correctness.
    \item Design normalization mechanisms for reducing length bias.
    \item Design different methods for classifying stylistic quality and content quality during the evaluation.
\end{itemize}

\section{Conclusion}
\label{sec:conclusion}
Large Language Models (LLMs) are increasingly used not only for generation tasks but also as automated evaluators through the emerging LLM-as-a-Judge (LaaJ) paradigm. While this paradigm enables scalable and efficient evaluation processes, it also introduces new security and reliability concerns that may compromise the integrity of automated assessment pipelines. Despite the growing adoption of LaaJ systems across multiple domains, the security implications associated with their use remain only partially explored.

To address this gap, this article provides a comprehensive survey of the security aspects related to LLM-as-a-Judge systems. We systematically reviewed the recent literature and analyzed the main threats and opportunities associated with the use of LLM-based judges in security-sensitive contexts. In particular, we examined several categories of attacks targeting or exploiting LaaJ systems, including prompt injection, backdoor attacks, bias exploitation, and manipulation of evaluation processes. Building on this analysis, we proposed a taxonomy that organizes the current literature according to five main perspectives: {\em(i)} attacks targeting LaaJ systems, {\em(ii)} attacks performed through LaaJ systems, {\em(iii)} defenses that leverage LaaJ as a security analysis or detection tool, {\em(iv)} LaaJ system as target of evaluation; and {\em(v)} the use of LaaJ as an evaluation strategy in security-related applications.

\begin{figure}
    \centering
    \includegraphics[width=0.6\linewidth]{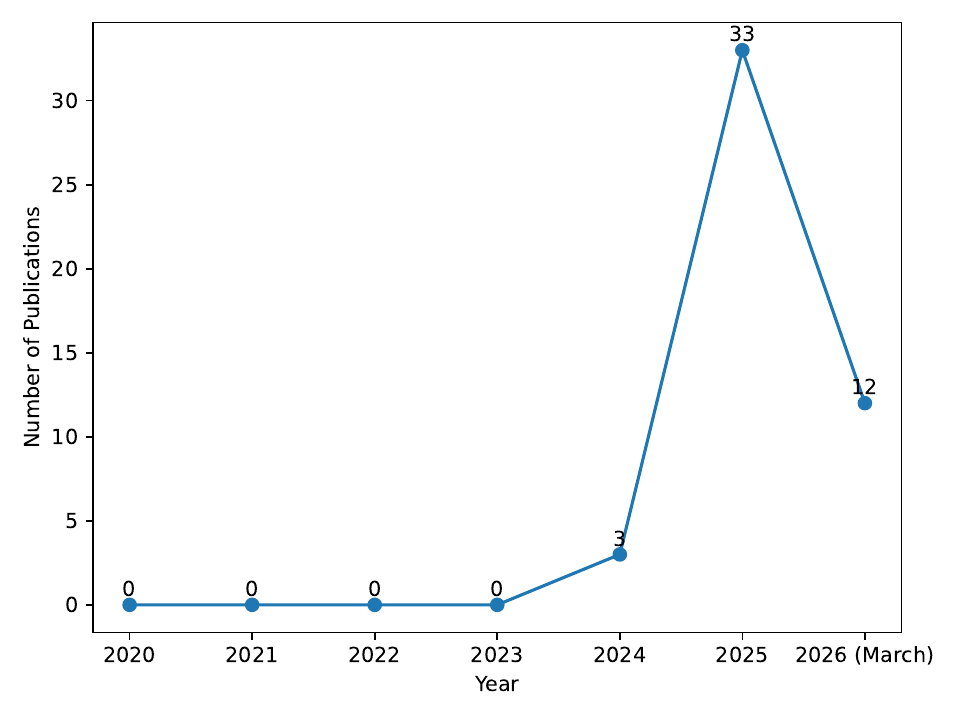}
    \caption{Yearly Distribution of Publications}
    \label{fig:pubYear}
\end{figure}

\begin{figure}
    \centering
    \includegraphics[width=0.6\linewidth]{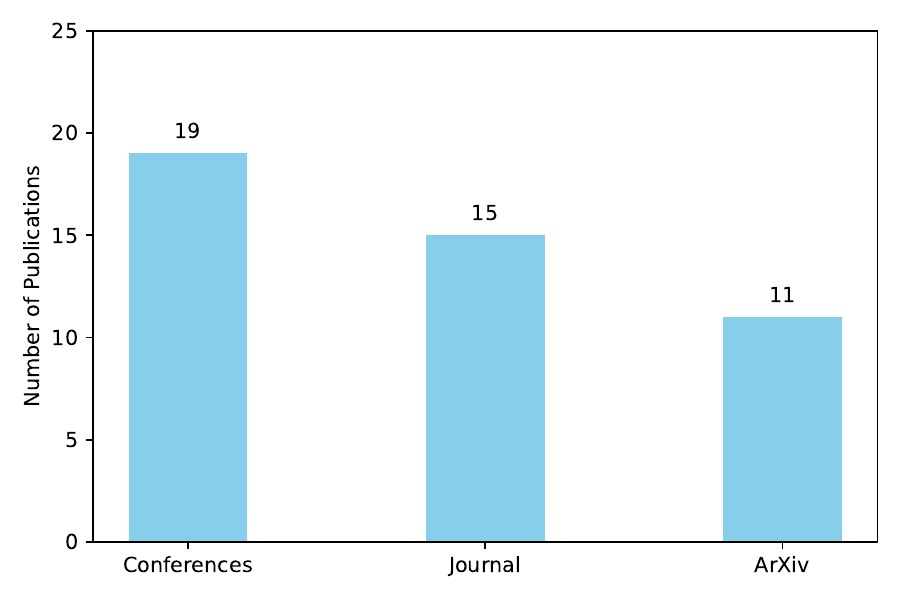}
    \caption{Number of Publications by Venue}
    \label{fig:pubVenue}
\end{figure}

The analysis of the publication trends further highlights the novelty and rapid evolution of this research area. The line chart in Figure \ref{fig:pubYear} shows the evolution of publications related to LLM-as-a-Judge security from 2024 to March 2026. In particular, there is a sharp increase in publications from 2024 (3 papers) to 2025 (33 papers), indicating a rapidly growing interest in the field. The trend observed in early 2026 already suggests a continued expansion of research activity in this domain. In addition, the distribution of the selected papers across venues provides further insights into the maturity of the field. As shown in the bar chart in Figure \ref{fig:pubVenue}, most studies appear in conferences, reflecting the fast-moving and exploratory nature of the research area. At the same time, ArXiv preprints represent a notable fraction of the literature, further underscoring the novelty of the topic and the tendency of researchers to disseminate early findings prior to formal publication.

Overall, our study highlights several limitations of existing approaches, including the vulnerability of LLM-based judges to adversarial manipulation, the lack of standardized evaluation methodologies, and the limited availability of robust benchmarks for assessing the security and reliability of LaaJ frameworks. These findings reveal significant research gaps and emphasize the need for more robust, transparent, and trustworthy evaluation mechanisms based on LLMs.

The works reviewed in this survey provide a foundation for further research on secure and reliable LLM-as-a-Judge systems. Future research directions include the development of standardized benchmarks for evaluating the robustness of LaaJ frameworks, the design of defenses against adversarial manipulation of LLM-based judges, and the exploration of hybrid evaluation strategies that combine automated judges with human oversight. Addressing these challenges will be crucial to ensure the safe and trustworthy deployment of LLM-as-a-Judge systems in real-world applications.

\bibliographystyle{plain}
\bibliography{biblio}
\end{document}